\overfullrule=0pt
\input harvmac
\input epsf.sty

\def\a{{\alpha}}

\def\l{{\lambda}}

\def\b{{\beta}}

\def\g{{\gamma}}

\def\d{{\delta}}

\def\s{{\sigma}}

\def\half{{1\over 2}}
\def\p{{\partial}}

\def\t{{\theta}}

\def\({\left(}
\def\){\right)}
\def\cF{{\cal F}}
\def\cI{{\cal I}}

%%%%%%%%%%%%%%%%%%%%%%%%%%%%%%%%%%%%%%%%%%%%%%%%%%%%%%%%%%%%%%%%
%%%%%   Dirac-Slash
%%%%%%%%%%%%%%%%%%%%%%%%%%%%%%%%%%%%%%%%%%%%%%%%%%%%%%%%%%%%%%%%
\def\slashchar#1{\setbox0=\hbox{$#1$}           % set a box for #1
   \dimen0=\wd0                                 % and get its size
   \setbox1=\hbox{/} \dimen1=\wd1               % get size of /
   \ifdim\dimen0>\dimen1                        % #1 is bigger
      \rlap{\hbox to \dimen0{\hfil/\hfil}}      % so center / in box
      #1                                        % and print #1
   \else                                        % / is bigger
      \rlap{\hbox to \dimen1{\hfil$#1$\hfil}}   % so center #1
      /                                         % and print /
   \fi}

%\draft
%%%%%%%%%%%%%%%%%%%%%%%%%%%%%%%%%%%%%%%%%%%%%%%%%%%%%%%%%%%%%%%%%%
%%%%%%%%%%%%%%%%%   Stuff for Figures  %%%%%%%%%%%%%%%%%%%%%%%%%%%
%%%%%%%%%%%%%%%%%%%%%%%%%%%%%%%%%%%%%%%%%%%%%%%%%%%%%%%%%%%%%%%%%%

\input epsf
\def\figin{\epsfcheck\figin}\def\figins{\epsfcheck\figins}
\def\epsfcheck{\ifx\epsfbox\UnDeFiNeD
\message{(NO epsf.tex, FIGURES WILL BE IGNORED)}
\gdef\figin##1{\vskip2in}\gdef\figins##1{\hskip.5in}% blank space instead
\else\message{(FIGURES WILL BE INCLUDED)}%
\gdef\figin##1{##1}\gdef\figins##1{##1}\fi}
\def\DefWarn#1{}
\def\figinsert{\goodbreak\midinsert}
\def\ifig#1#2#3{\DefWarn#1\xdef#1{Fig.~\the\figno}
\writedef{#1\leftbracket fig.\noexpand~\the\figno}%
\figinsert\figin{\centerline{#3}}\medskip\centerline{\vbox{\baselineskip12pt
\advance\hsize by -1truein\noindent\footnotefont\centerline{{\bf
Fig.~\the\figno}\ #2}}}
\bigskip\endinsert\global\advance\figno by1}
%%%%%%%%%%%%%%%%%%%%%%%%%%%%%%%%%%%%%%%%%%%%%%%%%%%%%%%%%

%%%%%%%%%%%%%%%%%%%%%%%%%%%%%%%%%%%%%%%%%%%%%%%%%%%%%%%%%%%%%%%%%%
\Title{\vbox{\rightline{AEI--2010--160}\rightline{MPP--2010--124}
\rightline{LMU--ASC--81/10}\vskip-1.75cm}}{\vbox{\centerline{
Six Open String Disk Amplitude in Pure Spinor Superspace}}}
\centerline{Carlos R. Mafra$^a$,
Oliver Schlotterer$^b$,
Stephan Stieberger$^b$, and
Dimitrios Tsimpis$^c$\foot{Address after November 1st: 
D\'epartement de Physique, Institut de Physique Nucl\'eaire de Lyon,
Universit\'e Claude Bernard Lyon 1.}}
\bigskip\medskip
\centerline{\it $^a$ Max--Planck--Institut f\"ur Gravitationsphysik} 
\centerline{\it Albert--Einstein--Institut}
\centerline{\it 14476 Potsdam, Germany}
\vskip4pt
\centerline{\it $^b$ Max--Planck--Institut f\"ur Physik}
\centerline{\it Werner--Heisenberg--Institut}
\centerline{\it 80805 M\"unchen, Germany}
\vskip4pt
\centerline{\it $^c$ Arnold--Sommerfeld--Center for Theoretical Physics,}
\centerline{\it  Department f\"ur Physik, Ludwig--Maximilians--Universit\"at 
M\"unchen}
\centerline{\it 80333 M\"unchen, Germany}
\smallskip
\centerline{\tt E-mails: crmafra@aei.mpg.de, olivers@mppmu.mpg.de,}
\centerline{\tt stephan.stieberger@mpp.mpg.de, dimitrios.tsimpis@lmu.de}

\bigskip\medskip
\centerline{\bf Abstract}
\vskip1pt
\noindent
The tree--level amplitude of six massless open strings is computed using the pure spinor formalism.
The OPE poles among integrated and unintegrated vertices can be efficiently organized
according to the cohomology of pure spinor superspace. The identification and use of these BRST structures
and their interplay with the system of equations fulfilled by the generalized Euler integrals
allow the full supersymmetric six--point amplitude to be written in compact form.
Furthermore, the complete set of extended Bern--Carrasco--Johansson relations are derived
from the monodromy  properties of the disk world--sheet
and explicitly verified for the supersymmetric numerator factors.

\Date{}
\noindent

\goodbreak

\lref\BernDW{
  Z.~Bern, L.J.~Dixon and D.A.~Kosower,
``On-Shell Methods in Perturbative QCD,''
  Annals Phys.\  {\bf 322}, 1587 (2007)
  [arXiv:0704.2798 [hep-ph]].
  %%CITATION = APNYA,322,1587;%%
}

\lref\BernFY{
  Z.~Bern, J.J.M.~Carrasco and H.~Johansson,
``The Structure of Multiloop Amplitudes in Gauge and Gravity Theories,''
  arXiv:1007.4297 [hep-th].
  %%CITATION = ARXIV:1007.4297;%%
}

\lref\KawaiXQ{
  H.~Kawai, D.C.~Lewellen and S.H.H.~Tye,
``A Relation Between Tree Amplitudes Of Closed And Open Strings,''
  Nucl.\ Phys.\  B {\bf 269}, 1 (1986).
  %%CITATION = NUPHA,B269,1;%%
}

\lref\BCJ{
  Z.~Bern, J.J.~M.~Carrasco and H.~Johansson,
  ``New Relations for Gauge-Theory Amplitudes,''
  Phys.\ Rev.\  D {\bf 78}, 085011 (2008)
  [arXiv:0805.3993 [hep-ph]].
  %%CITATION = PHRVA,D78,085011;%%
}

\lref\mafrabcj{
  C.R.~Mafra,
``Simplifying the Tree-level Superstring Massless Five-point Amplitude,''
  JHEP {\bf 1001}, 007 (2010)
  [arXiv:0909.5206 [hep-th]].
  %%CITATION = JHEPA,1001,007;%%
}

\lref\BjerrumBohrRD{
  N.E.J.~Bjerrum-Bohr, P.H.~Damgaard and P.~Vanhove,
  ``Minimal Basis for Gauge Theory Amplitudes,''
  Phys.\ Rev.\ Lett.\  {\bf 103}, 161602 (2009)
  [arXiv:0907.1425 [hep-th]].
  %%CITATION = PRLTA,103,161602;%%
}
\lref\ictp{
  N.~Berkovits,
  ``ICTP lectures on covariant quantization of the superstring,''
  arXiv:hep-th/0209059.
  %%CITATION = HEP-TH/0209059;%%
}
\lref\stieMultiGluon{
  S.~Stieberger and T.R.~Taylor,
``Amplitude for N-gluon superstring scattering,''
  Phys.\ Rev.\ Lett.\  {\bf 97}, 211601 (2006)
  [arXiv:hep-th/0607184];
  %%CITATION = PRLTA,97,211601;%%
  ``Multi-gluon scattering in open superstring theory,''
  Phys.\ Rev.\  D {\bf 74}, 126007 (2006)
  [arXiv:hep-th/0609175].
  %%CITATION = PHRVA,D74,126007;%%
}
\lref\stieS{
  S.~Stieberger and T.R.~Taylor,
  ``Supersymmetry Relations and MHV Amplitudes in Superstring Theory,''
  Nucl.\ Phys.\  B {\bf 793}, 83 (2008)
  [arXiv:0708.0574 [hep-th]].
  %%CITATION = NUPHA,B793,83;%%
}

\lref\StiebergerAM{
  S.~Stieberger and T.R.~Taylor,
 ``Complete Six--Gluon Disk Amplitude in Superstring Theory,''
  Nucl.\ Phys.\  B {\bf 801}, 128 (2008)
  [arXiv:0711.4354 [hep-th]].
  %%CITATION = NUPHA,B801,128;%%
}

\lref\PSSids{
  C.R.~Mafra,
  ``Pure Spinor Superspace Identities for Massless Four-point Kinematic
  Factors,''
  JHEP {\bf 0804}, 093 (2008)
  [arXiv:0801.0580 [hep-th]].
  %%CITATION = JHEPA,0804,093;%%
}

\lref\PSsuperspace{
  N.~Berkovits,
  ``Explaining pure spinor superspace,''
  arXiv:hep-th/0612021.
  %%CITATION = HEP-TH/0612021;%%
}

\lref\wittentwistor{
  E.~Witten,
  ``Twistor - Like Transform In Ten-Dimensions,''
  Nucl.\ Phys.\  B {\bf 266}, 245 (1986).
  %%CITATION = NUPHA,B266,245;%%
}
\lref\psf{
  N.~Berkovits,
  ``Super-Poincare covariant quantization of the superstring,''
  JHEP {\bf 0004}, 018 (2000)
  [arXiv:hep-th/0001035].
  %%CITATION = JHEPA,0004,018;%%
}
\lref\topological{
  N.~Berkovits,
  ``Pure spinor formalism as an N = 2 topological string,''
  JHEP {\bf 0510}, 089 (2005)
  [arXiv:hep-th/0509120].
  %%CITATION = JHEPA,0510,089;%%
}

\lref\anomaly{
  N.~Berkovits and C.R.~Mafra,
  ``Some superstring amplitude computations with the non-minimal pure spinor
  formalism,''
  JHEP {\bf 0611}, 079 (2006)
  [arXiv:hep-th/0607187].
  %%CITATION = JHEPA,0611,079;%%
}
\lref\FORM{
  J.A.M.~Vermaseren,
  ``New features of FORM,''
  arXiv:math-ph/0010025.
  %%CITATION = MATH-PH/0010025;%%
\semi
  M.~Tentyukov and J.A.M.~Vermaseren,
  ``The multithreaded version of FORM,''
  arXiv:hep-ph/0702279.
  %%CITATION = HEP-PH/0702279;%%
}
\lref\PSS{
  C.R.~Mafra,
  ``PSS: A FORM Program to Evaluate Pure Spinor Superspace Expressions,''
  arXiv:1007.4999 [hep-th].
  %%CITATION = ARXIV:1007.4999;%%
}
\lref\treelevel{
  N.~Berkovits and B.C.~Vallilo,
  ``Consistency of super-Poincare covariant superstring tree amplitudes,''
  JHEP {\bf 0007}, 015 (2000)
  [arXiv:hep-th/0004171].
  %%CITATION = JHEPA,0007,015;%%
}

\lref\StieOpr{
  D.~Oprisa and S.~Stieberger,
  ``Six gluon open superstring disk amplitude, multiple hypergeometric  series
  and Euler-Zagier sums,''
  arXiv:hep-th/0509042.
  %%CITATION = HEP-TH/0509042;%%
}
\lref\FTAmps{
  C.R.~Mafra,
  ``Towards Field Theory Amplitudes From the Cohomology of Pure Spinor
  Superspace,''
  JHEP {\bf 1011}, 096 (2010)
  [arXiv:1007.3639 [hep-th]].
  %%CITATION = JHEPA,1011,096;%%
}
\lref\FTnpt{
C.R.~Mafra, O.~Schlotterer, S.~Stieberger and D.~Tsimpis,
  ``A recursive formula for N-point SYM tree amplitudes,''
  arXiv:1012.3981 [hep-th].
  %%CITATION = ARXIV:1012.3981;%%
}
\lref\ggi{
  O.A.~Bedoya and N.~Berkovits,
  ``GGI Lectures on the Pure Spinor Formalism of the Superstring,''
  arXiv:0910.2254 [hep-th].
  %%CITATION = ARXIV:0910.2254;%%
}
\lref\thetaSYM{
  	J.P.~Harnad and S.~Shnider,
	``Constraints And Field Equations For Ten-Dimensional Superyang-Mills
  	Theory,''
  	Commun.\ Math.\ Phys.\  {\bf 106}, 183 (1986)
  	%%CITATION = CMPHA,106,183;%%
\semi
	H.~Ooguri, J.~Rahmfeld, H.~Robins and J.~Tannenhauser,
        ``Holography in superspace,''
        JHEP {\bf 0007}, 045 (2000)
        [arXiv:hep-th/0007104]
        %%CITATION = HEP-TH 0007104;%%
\semi
	P.A.~Grassi and L.~Tamassia,
        ``Vertex operators for closed superstrings,''
        JHEP {\bf 0407}, 071 (2004)
        [arXiv:hep-th/0405072].
        %%CITATION = HEP-TH 0405072;%%
}

\lref\vaman{
  D.~Vaman and Y.P.~Yao,
  ``Constraints and Generalized Gauge Transformations on Tree-Level Gluon and
  Graviton Amplitudes,''
  arXiv:1007.3475 [hep-th].
  %%CITATION = ARXIV:1007.3475;%%
}

\lref\tsimpis{
  G.~Policastro and D.~Tsimpis,
``$R^4$, purified,''
  Class.\ Quant.\ Grav.\  {\bf 23}, 4753 (2006)
  [arXiv:hep-th/0603165].
  %%CITATION = CQGRD,23,4753;%%
}
\lref\StiebergerHQ{
  S.~Stieberger,
  ``Open \& Closed vs. Pure Open String Disk Amplitudes,''
  arXiv:0907.2211 [hep-th].
  %%CITATION = ARXIV:0907.2211;%%
}
\lref\multiloop{
  N.~Berkovits,
  ``Multiloop amplitudes and vanishing theorems using the pure spinor
  formalism for the superstring,''
  JHEP {\bf 0409}, 047 (2004)
  [arXiv:hep-th/0406055].
  %%CITATION = JHEPA,0409,047;%%
}
\lref\twoloop{
  N.~Berkovits,
  ``Super-Poincare covariant two-loop superstring amplitudes,''
  JHEP {\bf 0601}, 005 (2006)
  [arXiv:hep-th/0503197].
  %%CITATION = JHEPA,0601,005;%%
}
\lref\regnmps{
  N.~Berkovits and N.~Nekrasov,
  ``Multiloop superstring amplitudes from non-minimal pure spinor formalism,''
  JHEP {\bf 0612}, 029 (2006)
  [arXiv:hep-th/0609012].
  %%CITATION = JHEPA,0612,029;%%
}
\lref\twolooptwo{
  N.~Berkovits and C.R.~Mafra,
  ``Equivalence of two-loop superstring amplitudes in the pure spinor and  RNS
  formalisms,''
  Phys.\ Rev.\ Lett.\  {\bf 96}, 011602 (2006)
  [arXiv:hep-th/0509234].
  %%CITATION = PRLTA,96,011602;%%
}
\lref\aisaka{
  Y.~Aisaka and N.~Berkovits,
  ``Pure Spinor Vertex Operators in Siegel Gauge and Loop Amplitude
  Regularization,''
  JHEP {\bf 0907}, 062 (2009)
  [arXiv:0903.3443 [hep-th]].
  %%CITATION = JHEPA,0907,062;%%
}
\lref\oneloopF{
  C.R.~Mafra,
  ``Four-point one-loop amplitude computation in the pure spinor formalism,''
  JHEP {\bf 0601}, 075 (2006)
  [arXiv:hep-th/0512052].
  %%CITATION = JHEPA,0601,075;%%
}
\lref\oneloopC{
  C.R.~Mafra and C.~Stahn,
  ``The One-loop Open Superstring Massless Five-point Amplitude with the
  Non-Minimal Pure Spinor Formalism,''
  JHEP {\bf 0903}, 126 (2009)
  [arXiv:0902.1539 [hep-th]].
  %%CITATION = JHEPA,0903,126;%%
}

\lref\tye{
  S.H.H. Tye and Y.~Zhang,
``Dual Identities inside the Gluon and the Graviton Scattering Amplitudes,''
  JHEP {\bf 1006}, 071 (2010)
  [arXiv:1003.1732 [hep-th]].
  %%CITATION = JHEPA,1006,071;%%
}

\lref\BjerrumBohrZS{
  N.E.J.~Bjerrum-Bohr, P.H.~Damgaard, T.~Sondergaard and P.~Vanhove,
``Monodromy and Jacobi-like Relations for Color-Ordered Amplitudes,''
  JHEP {\bf 1006}, 003 (2010)
  [arXiv:1003.2403 [hep-th]].
  %%CITATION = JHEPA,1006,003;%%
}

\lref\KK{
  R.~Kleiss and H.~Kuijf,
``Multigluon cross sections and 5-jet production at hadron colliders,''
  Nucl.\ Phys.\  B {\bf 312}, 616 (1989).
  %%CITATION = NUPHA,B312,616;%%
  }
\lref\KKi{
V.~Del Duca, L.J.~Dixon and F.~Maltoni,
``New color decompositions for gauge amplitudes at tree and loop level,''
  Nucl.\ Phys.\  B {\bf 571}, 51 (2000)
  [arXiv:hep-ph/9910563].
  %%CITATION = NUPHA,B571,51;%%
}
\lref\humberto{
  H.~Gomez,
  ``One-loop Superstring Amplitude From Integrals on Pure Spinors Space,''
  JHEP {\bf 0912}, 034 (2009)
  [arXiv:0910.3405 [hep-th]].
  %%CITATION = JHEPA,0912,034;%%
}
\lref\twoloopcoef{
  H.~Gomez and C.~R.~Mafra,
  ``The Overall Coefficient of the Two-loop Superstring Amplitude Using Pure
  Spinors,''
  JHEP {\bf 1005}, 017 (2010)
  [arXiv:1003.0678 [hep-th]].
  %%CITATION = JHEPA,1005,017;%%
}
\lref\Jaxo{D.~Binosi and L.~Theussl,
``JaxoDraw: A graphical user interface for drawing Feynman diagrams,''
  Comput.\ Phys.\ Commun.\  {\bf 161}, 76 (2004)
  [arXiv:hep-ph/0309015].
  %%CITATION = HEP-PH 0309015;%%
}
\lref\Medinas{
  R.~Medina, F.T.~Brandt and F.R.~Machado,
  ``The open superstring 5-point amplitude revisited,''
  JHEP {\bf 0207}, 071 (2002)
  [arXiv:hep-th/0208121]
  %%CITATION = JHEPA,0207,071;%%
\semi
  L.A.~Barreiro and R.~Medina,
  ``5-field terms in the open superstring effective action,''
  JHEP {\bf 0503}, 055 (2005)
  [arXiv:hep-th/0503182].
  %%CITATION = JHEPA,0503,055;%%
}

\listtoc
\writetoc
\break

%********************
\newsec{Introduction}
%********************

Elementary particle physics relies on scattering experiments.
The physical cross sections, determined by the scattering amplitudes, reflect the 
properties of underlying interactions. Already at the tree--level, such computations 
can be quite complicated, especially when a large number of external particles is 
involved, 
like in the scattering processes describing multi--jet production at hadron colliders. 
During the last years remarkable progress has been accumulated in our understanding and 
in our ability to compute scattering amplitudes, both for theoretical and 
phenomenological 
purposes, cf. Ref.~\BernDW\ for a recent account.

Scattering amplitudes in gauge and gravity theories have a remarkably rich 
yet simple structure, 
allowing  to develop even more powerful tools to understand their behavior.
Various relations within or between 
gravity and gauge theory scattering amplitudes suggest a unification within or 
between these 
theories of the sort inherent to string theory, cf. Ref.~\BernFY\ for a recent review.
Some field--theory properties of scattering amplitudes can be easily
derived and proven by string theory. One notorious example are the 
Kawai, Lewellen and Tye (KLT) relations, which express a graviton amplitude as a sum of 
squares of partial color ordered gluon amplitudes \KawaiXQ.
Another example are the so--called
Bern, Carrasco and Johansson (BCJ) relations, which relate
various partial color--ordered subamplitudes \BCJ. 
These relations have a natural explanation in 
string theory: they simply are consequences of monodromy properties on the 
string world--sheet \refs{\BjerrumBohrRD,\StiebergerHQ}.
Hence, world--sheet symmetries of string amplitudes turn out to have profound 
impact on the structure of field--theory amplitudes itself.
Therefore, the hidden structures and symmetries of superstring disk scattering 
amplitudes prove to be useful in revealing properties and symmetries of field--theory 
amplitudes.

Scattering amplitudes are also of  considerable theoretical interest 
in the framework of a full fledged superstring theory. 
The pure spinor formalism \refs{\psf,\topological} has been useful for quantizing 
the superstring in Ramond--Ramond backgrounds, and has considerably simplified the 
computation of multi--loop superstring scattering 
amplitudes \refs{\multiloop\twoloop\regnmps\aisaka\twolooptwo\oneloopF\PSSids
\oneloopC\humberto-\twoloopcoef}. 
Four-- and five--point tree--level \treelevel\ amplitudes have been computed in pure spinor 
superspace and cast into compact form\foot{In the RNS formalism the five--gluon amplitude has been computed in Refs. \refs{\Medinas,\StieOpr}, 
while the six--gluon amplitude has been computed in Refs. \refs{\StieOpr\stieMultiGluon\stieS-\StiebergerAM}.} in Refs. \refs{\PSSids,\mafrabcj}

The pure spinor formalism  might  also be used to describe $D=11$ 
supergravity and M--theory.
The BRST cohomology properties of the pure 
spinor superspace \PSsuperspace\ are useful not only to simplify the string 
amplitudes \refs{\PSSids,\mafrabcj}
but also has recently been suggested of allowing field--theory amplitudes to be 
obtained directly \refs{\FTAmps,\FTnpt}.
Furthermore, the BRST cohomology sheds light on the structure and organization of 
the terms in higher--point open superstring amplitudes.
Hence, it is of fundamental importance to pursue multi--leg amplitude computations in
pure spinor superspace and anticipate their underlying symmetries, e.g. 
by exploiting their  BRST cohomology properties.

In this work we show that the color--ordered open superstring six--point disk 
amplitude computed with the pure spinor formalism is given by
\eqn\simpleSpt{\eqalign{
A(1,2,3,4,5,6) &=
\langle T_{54}T_{32}T_{16}\rangle F_1
+ \langle T_{52}T_{43}T_{16}\rangle F_2
+ \langle T_{53}T_{42}T_{16}\rangle F_3\cr
&+ \langle T_{123}E_{456}\rangle F_4
+ \langle T_{324}E_{561}\rangle F_5
+ \langle T_{435}E_{216}\rangle F_6
+ \langle T_{325}E_{416}\rangle F_7\cr
&+ \langle T_{124}E_{356}\rangle F_8
+ \langle T_{352}E_{416}\rangle F_9
+ \langle T_{241}E_{356}\rangle F_{10}
+ (1\leftrightarrow 5,\, 2\leftrightarrow 4).
}}
The pure spinor bracket $\langle {\ldots} \rangle$ was defined in
\psf\ and selects the terms proportional to 
$(\l\g^m\t)(\l\g^n\t)(\l\g^p\t) (\t\g_{mnp}\t)$, which is the unique
element in the cohomology of the BRST charge at ghost number three.
The factors of $F_i$ (and their image $F'_i$ under 
$1\leftrightarrow 5, 2\leftrightarrow 4$)
denote certain combinations of generalized Euler integrals
whose momentum expansions obtained with the
methods of \StieOpr\ are listed in Appendix B.
The factors of $[T_{ij}, T_{ijk}, E_{ijk}]$ are
pure spinor BRST building blocks whose definitions and properties will
be explained in section 2.
The result \simpleSpt\ simplifies and extends the RNS computations of \StieOpr\ to the full supermultiplet.
A compact expression for the full six--point superstring amplitude
in the $D=4$ helicity basis can be found in 
\refs{\stieMultiGluon\stieS-\StiebergerAM}.
Using the FORM \FORM\ program described in \PSS, the six--gluon
component expansion of \simpleSpt\ can be extracted. In fact, up to the order 
$\alpha'^3$, which is available from the authors upon request, we have explicitly verified that the latter agrees with the result of \StieOpr. 
Furthermore, the field--theory limit $\a'\to 0$ of \simpleSpt\ is:
\eqn\FTlimitintro{\eqalign{
A_{\rm SYM}(1,2,3,4,5,6) &=
 {\langle T_{12}T_{34}T_{56}\rangle \over 3\, s_1 s_3 s_5}
 + \half \langle \( {T_{123}\over s_1 t_1} - {T_{231}\over s_2 t_1}\)  E_{456}\rangle 
 + {\rm cyclic}(1{\ldots} 6)\cr
&= {\langle T_{12}T_{34}T_{56}\rangle \over s_1 s_3 s_5}
+ {\langle T_{54}T_{32}T_{16}\rangle\over s_2 s_4 s_6}
+ {\langle T_{123}E_{456}\rangle\over s_1 t_1}
+ {\langle T_{543}E_{216}\rangle\over s_4 t_3}\cr
&+ {\langle T_{321}E_{456}\rangle \over s_2 t_1}
+ {\langle T_{345}E_{216}\rangle \over s_3 t_3}
+ {\langle T_{432}E_{561}\rangle \over s_3 t_2}
+ {\langle T_{234}E_{561}\rangle \over s_2 t_2}\ .}}
The expression \FTlimitintro\ agrees with the superspace expression recently proposed based on BRST cohomology  \FTAmps. The superspace expressions
in \FTlimitintro\ can be interpreted \FTAmps\ in terms of Feynman diagrams which use only
cubic vertices as discussed in \BCJ. The diagrams associated with the
three terms in the first line of \FTlimitintro\ --- which generate
the full amplitude upon cyclic symmetrization --- are depicted in Figure 1.
\ifig\figi{The five field theory 
diagrams and their corresponding pure spinor superspace expressions.}
{\epsfxsize=0.95\hsize\epsfbox{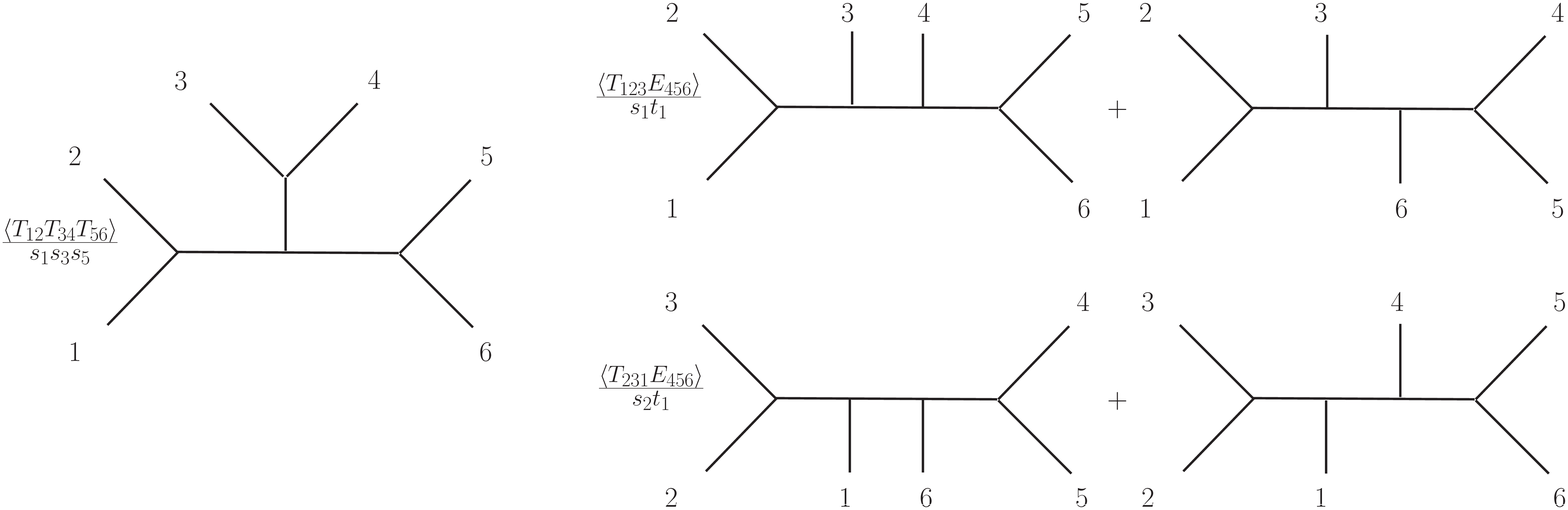}}

%********************************
\newsec{Pure spinor preliminaries}
%********************************

The prescription to compute the massless open superstring 
six--point tree-level amplitude is given by \psf
\eqn\spt{
{\cal A}(1,2,3,4,5,6) =
\langle V^1(z_1)V^5(z_5)V^6(z_6) \int dz_2 U^2(z_2) \int dz_3 U^3(z_3) \int dz_4 U^4(z_4)\rangle,
}
where $V^i(z_i)$ and $U^i(z_i)$ are the vertex operators with conformal weight zero and one,
\eqn\vertices{
V^i(z_i) = \l^\a(z_i) A^i_\a(X^i,\t), \quad 
U^i(z_i) = \p\t^\a A^i_\a + \Pi^m A^i_m + d_\a W^\a_i + \half {\cal F}_{mn}^i N^{mn},
}
and the positions of the unintegrated vertices are fixed by $SL(2,{\bf R})$ invariance to arbitrary locations, 
which in this paper are chosen as $(z_1,z_5,z_6) = (0,1,\infty)$.
The amplitude \spt\ represents the color ordered subamplitude, given by
${\cal A}(1,2,3,4,5,6)=\Tr(T^{a_1}T^{a_2}T^{a_3}T^{a_4}T^{a_5}T^{a_6})\ 
A(1,2,3,4,5,6)$, with the Chan--Paton factors in the adjoint representation.

The operators $[d_\a,\Pi^m,N^{mn},\p\t^\a, \l^\a, w_\a]$ satisfy the OPEs\foot{For reviews of the pure
spinor formalism, see \refs{\ictp,\ggi}.}
$$
d_\a(z) V(w) \rightarrow  {D_\a V(w)\over z- w}, \quad
\Pi^m(z) V(w) \rightarrow -  {k^m V(w)\over z- w}, \quad
d_\a(z) \Pi^m(w) \rightarrow  {(\g^m\p\t)_\a\over z- w}
$$
$$
d_\a(z)d_\b(w) \rightarrow -  {\g^m_{\a\b}\Pi_m\over z- w}, \quad
\Pi^m(z)\Pi^n(w) \rightarrow -  {\eta^{mn}\over (z - w)^2}, \quad
d_\a(z)\t^\b(w) \rightarrow  {\d^\b_\a\over (z- w)}
$$
$$
d_\a(z)\p\t^\b(w) \rightarrow  {\d^\b_\a\over (z- w)^2},\quad
w(z)_\a\l^\b(w) \rightarrow - {\d^\b_\a\over z - w}, \quad
N^{mn}(z)\l^\a(w) \rightarrow - \half {(\l\g^{mn})^\a\over z- w}
$$
$$
N^{mn}(z)N_{pq}(w) \rightarrow + {4 \over z - w}N^{[m}_{\phantom{m}[p}\d^{n]}_{q]}
- {6 \over (z - w)^2}\d^n_{[p}\d^m_{q]}
$$
where $V(w)$ is an arbitrary conformal weight-zero superfield, $N^{mn} = \half (\l\g^{mn}w)$ are
the pure spinor Lorentz currents and the antisymmetrization bracket $[{\ldots}]$ encompassing $N$ indices 
is defined to contain an
overall $1/N!$. The super-Yang-Mills superfields $[A_\a,A_m,W^\a,{\cal F}_{mn}]$
satisfy the equations of motion \refs{\wittentwistor,\ictp}
$$
D_\a A_\b + D_\b A_\a = \g^m_{\a\b} A_m,\quad
D_\a A_m = (\g_m W)_\a + k_m A_\a,
$$
\eqn\SYM{
D_\a{\cal F}_{mn} = 2k_{[m} (\g_{n]} W)_\a, \quad
D_\a W^{\b} = {1\over 4}(\g^{mn})^{\phantom{m}\b}_\a{\cal F}_{mn},
}
and have the following $\t$--expansions in the gauge $\t^\a A_\a =0$ \refs{\thetaSYM,\tsimpis},
$$
A_{\a}(x,\t)={1\over 2}a_m(\g^m\t)_\a -{1\over 3}(\xi\g_m\t)(\g^m\t)_\a
-{1\over 32}F_{mn}(\g_p\t)_\a (\t\g^{mnp}\t) + \ldots
$$
$$
A_{m}(x,\t) = a_m - (\xi\g_m\t) - {1\over 8}(\t\g_m\g^{pq}\t)F_{pq}
         + {1\over 12}(\t\g_m\g^{pq}\t)(\p_p\xi\g_q\t) + \ldots
$$
$$
W^{\a}(x,\t) = \xi^{\a} - {1\over 4}(\g^{mn}\t)^{\a} F_{mn}
           + {1\over 4}(\g^{mn}\t)^{\a}(\p_m\xi\g_n\t)
	   + {1\over 48}(\g^{mn}\t)^{\a}(\t\g_n\g^{pq}\t)\p_m F_{pq} 
	   + \ldots
$$
\eqn\expansions{
\cF_{mn}(x,\t) = F_{mn} - 2(\p_{[m}\xi\g_{n]}\t) + {1\over
4}(\t\g_{[m}\g^{pq}\t)\p_{n]}F_{pq} + {\ldots},
}
where $a_m(x) = e_m {\rm e}^{ik\cdot x}$, $\xi^{\a}(x) =\chi^\a {\rm e}^{ik\cdot x}$ are the bosonic
and fermionic polarizations and $F_{mn} = 2\p_{[m} a_{n]}$ is the field-strength.

%***************************************
\subsec{Six--point kinematic invariants}
%***************************************

By using momentum conservation an $N$--point amplitude can be written in terms of $N(N-3)/3$ kinematic
invariants. It is convenient to define \stieMultiGluon
$$
s_{ij}= 2 \alpha' k_i\cdot k_j,\quad s_i = \alpha' (k_i + k_{i+1})^2,\quad
s_{ijk}= s_{ij} + s_{ik} + s_{jk},
$$
\eqn\Sinv{
t_l=\alpha'  (k_l + k_{l+1} + k_{l+2})^2,\quad i,j,k=1,{\ldots},6,\quad l=1,2,3
}
such that all six--point kinematic invariants can be written\foot{From now on we set $\a'=1/2$ unless otherwise stated.}
in terms of $s_1,{\ldots},s_6$ and $t_1, t_2, t_3$,
\eqn\kinsol{\eqalign{
s_{13} &=  - s_1  -  s_2  +  t_1, \quad
s_{14} =  s_2  +  s_5  -  t_1  -  t_2,\quad
s_{15} =  -  s_5  -  s_6  +  t_2\ ,\cr
s_{24} &=  -  s_2  -  s_3  +  t_2,\quad
s_{25} =    s_3  +  s_6  -  t_2  -  t_3,\quad
s_{26} =  -  s_1  -  s_6  +  t_3\ ,\cr
s_{35} &=  -  s_3  -  s_4  +  t_3,\quad
s_{36} =   s_1  +  s_4  -  t_1  -  t_3,\quad
s_{46} =  -  s_4  -  s_5  +  t_1\ .}}
The $s_i$ and $t_i$ variables have well-defined transformations under cyclic transformations; 
$s_{i+6}=s_i$ and $t_{i+3} = t_i$. Furthermore, under the worldsheet
parity transformation $1\leftrightarrow 5$, $2\leftrightarrow 4$ (also known as twist) the kinematic invariants $s_i$ and $t_i$  transform as
\eqn\parity{
s_1 \leftrightarrow s_4, \quad s_2 \leftrightarrow s_3,\quad s_5 \leftrightarrow s_6,\quad
t_1 \leftrightarrow t_3, \quad t_2 \leftrightarrow t_2\ .}
The six--point amplitude can be shown to be invariant under worldsheet
parity, i.e.: $A(1,2,3,4,5,6) = A(6,5,4,3,2,1)$.

%****************************
\subsec{BRST building blocks}
%****************************

The simple form of the BRST charge in the pure spinor formalism when 
acting on superfields, $Q=\l^\a D_\a$, turns out to allow an efficient
method to organize the computations. Inspired by the explicit superspace
computations of the four- and five-point amplitudes in \refs{\PSSids,\mafrabcj},
this method to handle the computations efficiently consists in identifying the
so-called {\it BRST building blocks} which transform covariantly under
the pure spinor BRST charge.

Using this BRST-covariant organization together with
the kinematic pole expansion of the tree-level amplitudes discussed in \BCJ,
Ans\"atze for the six-- and seven--point super-Yang-Mills amplitudes were
presented in \FTAmps. In this section more BRST building blocks which appear naturally
in the full superstring six--point amplitude will be identified, expanding and improving
the set used in \FTAmps.

The OPEs can be used to define $L_{ij}$, $L_{jiki}$ and $L_{jikili}$ as
\eqn\Ldefs{
V^i(z_i) U^j(z_j) \rightarrow {L_{ij}\over z_{ij}}, \quad
L_{ij}(z_i)U^k(z_k) \rightarrow {L_{jiki}\over z_{ik}},\quad
L_{jiki}(z_i)U^l(z_l) \rightarrow {L_{jikili}\over z_{li}}.
}
Their explicit expressions are written in Appendix A.
It is also convenient to define
\eqn\DR{
D_{ij} = (A^i\cdot A^j), \quad R_{ijk} = D_{ij}(A^k\cdot(k^i+k^j)),
}
\eqn\Oijk{
O_{ijk} = {1\over 2}D_{ij}(k^i \cdot A^k)  + {\rm antisymmetrization \ in} \ i,j,k ,
}
which are motivated by the residues of $U^i(z_i)U^j(z_j)$ and $U^i(z_i)U^j(z_j)U^k(z_k)$
double poles.

Computing the BRST variations of \Ldefs\ one gets \FTAmps,
\eqn\QLij{
QL_{ij} = - s_{ij} V_i V_j,
}
\eqn\BRSTLjiki{
Q L_{jiki} = s_{ij}\big[ L_{jk} V_i - L_{ik}V_j + L_{ij}V_k\big]
- s_{ijk} L_{ij}V_k,
}
\eqn\BRSTLjikili{
QL_{jikili} = 
  (k_i \cdot k_j) \(V_i  L_{kjlj} + L_{kili} V_j\)
+ \big[(k_i + k_j) \cdot k_k \big] L_{jili} V_k
}
$$
+ \big[(k_i + k_j + k_k) \cdot k_l \big]  L_{jiki} V_l
+ (k_i \cdot k_j)  \big[  L_{ik}  L_{jl}  +  L_{il} L_{jk}\big]
+  \big[(k_i + k_j) \cdot k_k \big] L_{ij} L_{kl}.
$$
Using the SYM equations of motion it is easy to see that the symmetric piece
of $L_{ij}$ is BRST-exact,
\eqn\brstrivial{
QD_{ij} = L_{ij} + L_{ji}
}
which suggests defining the antisymmetric part to be the first 
composite BRST building block $T_{ij}$,
\eqn\Tij{
T_{ij} \equiv L_{ij} - \half QD_{ij}, \quad QT_{ij} = - s_{ij}V_i V_j
}
Removing the BRST-exact part of $L_{ij}$ in the RHS of \BRSTLjiki\ leads
to the definition
\eqn\Ttijk{
{\tilde T}_{ijk} \equiv L_{jiki} - {s_{ij}\over 2}\big[
D_{jk}V_i - D_{ik}V_j + D_{ij}V_k\big] + {s_{ijk}\over 2}D_{ij}V_k.
}
Its BRST variation $Q {\tilde T}_{ijk} = s_{ij} T_{\{ij}V_{k\}}
- s_{ijk} T_{ij}V_k$ is written in terms of $T_{ij}$ instead
of $L_{ij}$, where $\{{\ldots} \}$ denotes the cyclic symmetrization
of the enclosed indices.
Note that ${\tilde T}_{(ij)k}$ is BRST-exact
\eqn\trivialT{
QR_{ijk} = {\tilde T}_{ijk} + {\tilde T}_{jik},
}
and therefore we extract the antisymmetric $[ij]$ part of ${\tilde T}_{ijk}$ as
\eqn\Tbarijk{
{\tilde T}_{[ij]k} \equiv {\tilde T}_{ijk} - \half QR_{ijk}.
}
Finally, using the definitions \Oijk\ and \Tbarijk\ it is possible to
show that $O_{ijk}$ obeys
\eqn\QOijk{
QO_{ijk} = - {\tilde T}_{[ij]k} - {\tilde T}_{[ki]j} - {\tilde T}_{[jk]i}
}
which finally suggests the definition of the next building block $T_{ijk}$,
\eqn\Tijk{
T_{ijk} \equiv {\tilde T}_{[ij]k} + {1\over 3}QO_{ijk} = 
{2\over 3}{\tilde T}_{[ij]k}
-{1\over 3}{\tilde T}_{[ki]j} 
-{1\over 3}{\tilde T}_{[jk]i},
}
which satisfies
\eqn\QTijk{
\quad Q T_{ijk} = s_{ij} T_{\{ij}V_{k\}} - s_{ijk} T_{ij}V_k.
}
Note from \Tijk\ that $T_{ijk}$ has the symmetries of a $(2,1)$-hook, 
\eqn\Tijkhooks{
T_{ijk} = T_{[ij]k}, \quad T_{\{ijk\}} = 0.
}
It is also  convenient to define 
\eqn\Eijk{
E_{ijk} \equiv {V_i T_{jk} \over s_{jk}}  + {T_{ij}V_k\over s_{ij}}
}
which is BRST-exact,
\eqn\Etrivial{
E_{ijk} = -{1\over s_{ijk}} Q\( {T_{ijk}\over s_{ij}} + {T_{kji}\over s_{kj}}\), 
\quad QE_{ijk} = 0.
}
Following the same BRST reasoning, the next building block $\tilde T_{ijkl}$ is defined by
$$
\tilde T_{ijkl} \equiv L_{jikili}
- \half\big[
       (s_{ijk} - s_{ij})(T_{kl}D_{ij} - T_{ij}D_{kl})
       + s_{ij}\(T_{jk} D_{il} + T_{jl}D_{ik} - T_{ik}D_{jl} - T_{il}D_{jk}\)
       \big]
$$
$$
       - {1\over 4} \big[
        s_{ij}(D_{ik}(L_{jl} + L_{lj}) + D_{il}(L_{jk} + L_{kj}))
       - (s_{ij} - s_{ijk})D_{ij}(L_{kl} + L_{lk})
       \big]
$$
$$
       -\half \big[ s_{ijkl} R_{ijk}V_l
       + s_{ijk}(R_{ijl} V_k - R_{ijk}V_l)
       - s_{ij}( R_{jkl}V_i
       - R_{ikl}V_j
       + R_{ijl}V_k)
       \big]
$$
\eqn\Tijkl{
       + {1\over 3}\big[ s_{ijkl} O_{ijk} V_{l}
       + s_{ijk}( O_{ijl}V_{k}
                - O_{ijk}V_{l})
       - s_{ij}(O_{jkl}V_{i}
                - O_{ikl}V_{j}
                + O_{ijl}V_{k})
       \big],
}
which satisfies
\eqn\QTijkl{
Q\tilde T_{ijkl} = s_{ijkl}  T_{ijk}  V_l
+  s_{ijk}  \(  T_{ijl}  V_k  -  T_{ijk}  V_l  +  T_{ij}  T_{kl} \)
 + s_{ij}  \(  V_{\{i}  T_{jk\}l} -  T_{\{ij}  T_{k\}l}\).
}
The corrections containing $D_{ij}$ in the first two lines of \Tijkl\ are required to
make the BRST transformation $QT_{ijkl}$ 
be written in terms of $T_{ij}$ and ${\tilde T}_{ijk}$ rather than $L_{ij}$
and $L_{jiki}$.
Analogously, the $R_{ijk}$ corrections in the third line are needed to further rewrite
${\tilde T}_{ijk}$ in terms of ${\tilde T}_{[ij]k}$ and finally, the $O_{ijk}$ corrections
in the fourth line of \Tijkl\ allow ${\tilde T}_{[ij]k}$ to be rewritten in terms of the
building block $T_{ijk}$ of \Tijk. Therefore the RHS of the BRST variation \QTijkl\ is composed
only out of building blocks.

Using \QTijkl\ one can show that
\eqn\Tijklhooks{
Q \tilde T_{(ij)kl} = Q \tilde T_{[ijk]l} = Q( \tilde T_{ij[kl]} +  \tilde T_{kl[ij]}) = 0
}
and we expect that all these combinations are in fact BRST-exact. For example,
\eqn\nontrivial{
{\tilde T}_{ijkl} + {\tilde T}_{jikl} = QR_{ijkl}
}
where
\eqn\Rijkl{
R_{ijkl} = - R_{ijk}\ [(k^i+k^j+k^k)\cdot A^l] - {1\over 4}s_{ij}\big( D_{ik}D_{jl} + D_{jk}D_{il}\big)
}
Appropriate redefinitions $T_{ijkl} = \tilde T_{ijkl} + Q(\ldots)$ lead to a building block with four legs which obeys the symmetry properties \Tijklhooks \ by itself without $Q$ action:
\eqn\TTijklhooks{
T_{(ij)kl} =  T_{[ijk]l} =  T_{ij[kl]} +  T_{kl[ij]} = 0
}
Since $\tilde T_{ijkl}$ enters the six point amplitude in the combination $\langle \tilde T_{ijkl} V_m V_n \rangle = {-1 \over s_{mn}} \langle \tilde T_{ijkl} QT_{mn} \rangle$ only, the BRST exact parts decouple and we can replace $\tilde T_{ijkl} \leftrightarrow T_{ijkl}$ in all instances throughout this work.

%**********************************************************
\newsec{The six--point amplitude in pure spinor superspace}
%**********************************************************

With the conventions of the previous section, the following open-string 
six--point subamplitude will be computed
\eqn\sixpt{
{\cal A}(1,2,3,4,5,6) =
\langle V^1(z_1)V^5(z_5)V^6(z_6) \int dz_2 U^2(z_2) \int dz_3 U^3(z_3) \int dz_4 U^4(z_4)\rangle.
}
It would be reasonable to expect that the full explicit computation of this correlator becomes a rather
big and tedious expression to work with. However, we will show there are some simplifying features
of the pure spinor formalism which allow for an efficient
evaluation of \sixpt\ leading to a simple and compact result written in pure
spinor superspace.

To achieve this simplification, we exploit the interplay between kinematic factors in pure spinor superspace and their associated integrals. They
both obey different sets of identities which, when considered together, lead to
many cancellations at the superspace level.

Identities among the kinematic factors arise from amplitude's independence on the order in which the conformal weight-one variables are integrated out \mafrabcj.
As will become apparent below, an early application of this method 
reduces the amount of explicit superfield manipulations considerably. Furthermore, the pure spinor
computations are best organized using the BRST building-blocks of the previous section,
which has the additional benefit of reusing elements from amplitudes with a lower number of legs.

It is convenient to organize the six--point subamplitude in terms of all possible OPE contractions of the integrated vertex operators. Each OPE contribution is associated with its specific kinematic factor and $z_{ij}$ dependent denominator for the worldsheet integration, as in the 5-point amplitude of \mafrabcj.
Using the OPE's to eliminate the conformal weight-one variables with positions $(234)$ and
setting $\a' =1/2$,
the subamplitude $A(1,2,3,4,5,6)$
is written as the sum of 24 single- and ten double-pole integrals\foot{Single/double
pole integral denote the origin of the $z_i$ dependence, whether they come from the single or 
double poles in the OPE's. It is easy to see that the naive number of single-pole integrals
in a $N$--point open-string amplitude is $(N-2)!$.}
$$
%\qquad\quad\qquad\qquad 
A(1,2,3,4,5,6)
= \int dz_2 \int dz_3 \int dz_4 \prod_{i<j}^6 |z_{ij}|^{-s_{ij}}
$$
$$
\times\Big\{
  {\langle L_{213141} V_5 V_6\rangle \over z_{21}z_{31}z_{41}}
+ {\langle L_{213145} V_6 \rangle \over z_{21}z_{31}z_{45}}
+ {\langle L_{213541} V_6 \rangle \over z_{21}z_{35}z_{41}}
+ {\langle L_{213545} V_6 \rangle \over z_{21}z_{35}z_{45}}
+ {\langle L_{253141} V_6 \rangle \over z_{25}z_{31}z_{41}}
$$
$$
+ {\langle L_{253145} V_6 \rangle \over z_{25}z_{31}z_{45}}
+ {\langle L_{253541} V_6    \rangle \over z_{25}z_{35}z_{41}}
+ {\langle L_{253545} V_1 V_6\rangle \over z_{25}z_{35}z_{45}}
+ {\langle L_{213441} V_5 V_6\rangle \over z_{21}z_{34}z_{41}}
+ {\langle L_{213445} V_6    \rangle \over z_{21}z_{34}z_{45}}
$$
$$
+ {\langle L_{253441} V_6    \rangle \over z_{25}z_{34}z_{41}}
+ {\langle L_{253445} V_1 V_6\rangle \over z_{25}z_{34}z_{45}}
+ {\langle L_{243141} V_5 V_6\rangle \over z_{24}z_{31}z_{41}}
+ {\langle L_{243145} V_6    \rangle \over z_{24}z_{31}z_{45}}
+ {\langle L_{243541} V_6    \rangle \over z_{24}z_{35}z_{41}}
$$
\eqn\manyLs{\eqalign{
&+ {\langle L_{243545} V_1 V_6\rangle \over z_{24}z_{35}z_{45}}
+ {\langle L_{233141} V_5 V_6\rangle \over z_{23}z_{31}z_{41}}
+ {\langle L_{233541} V_6    \rangle \over z_{23}z_{35}z_{41}}
+ {\langle L_{233145} V_6 \rangle \over z_{23}z_{31}z_{45}}
+ {\langle L_{233545} V_1 V_6\rangle \over z_{23}z_{35}z_{45}}\cr
&+ {\langle L_{233441} V_5 V_6\rangle \over z_{23}z_{34}z_{41}}
+ {\langle L_{233445} V_1 V_6\rangle \over z_{23}z_{34}z_{45}}
+ {\langle L_{243431} V_5 V_6\rangle \over z_{24}z_{34}z_{31}}
+ {\langle L_{243435} V_1 V_6\rangle \over z_{24}z_{34}z_{35}}
+ {\langle L_{213434} V_5 V_6\rangle \over z_{21}   z_{34}^2}\cr
&+ {\langle L_{253434} V_1 V_6\rangle \over z_{25}   z_{34}^2}
+ {\langle L_{242431} V_5 V_6\rangle \over z_{24}^2 z_{31}}
+ {\langle L_{242435} V_1 V_6\rangle \over z_{24}^2 z_{35}}
+ {\langle L_{232341} V_5 V_6\rangle \over z_{23}^2 z_{41}}
+ {\langle L_{232345} V_1 V_6\rangle \over z_{23}^2 z_{45}}\cr
&+ {\langle L_{233434} V_1 V_5 V_6\rangle \over z_{23}   z_{34}^2}
+ {\langle L_{243434} V_1 V_5 V_6\rangle \over z_{24}   z_{34}^2}
+ {\langle L_{242434} V_1 V_5 V_6\rangle \over z_{24}^2 z_{34}}
+ {\langle L_{232334} V_1 V_5 V_6\rangle \over z_{23}^2 z_{34}}\Big\}\ .}}
The last ten double--pole integrals and their kinematic factors will be considered
separately below. 
Regarding the twenty four single-pole kinematic factors, fifteen can be obtained
by $1\leftrightarrow 5$, $2\leftrightarrow 4$ relabellings of
$$
\langle L_{213141} V_5 V_6\rangle,\; \langle L_{213145} V_6\rangle,\; \langle L_{213441}V_5V_6\rangle,\;
\langle L_{213445} V_6\rangle,\; \langle L_{213541} V_6\rangle,\;
$$
\eqn\setLs{
\langle L_{213545}V_6\rangle,\; \langle L_{233141}V_5 V_6\rangle,\;
\langle L_{233145}V_6\rangle,\; \langle L_{233441}V_5 V_6\rangle 
}
and it will now be shown that BCJ-like kinematic identities reduce the
number of independent kinematics to only four.

%*********************************************************
\subsec{Single pole integrands and BCJ-inspired technique}
%*********************************************************

The Bern--Carrasco--Johansson (BCJ) kinematic identities are relations
among the kinematic factors associated to different kinematic poles in the
field-theory scattering amplitudes \BCJ. In string theory, exploiting the 
independence of the CFT correlator on the order in which the OPE expansions are used one
obtains BCJ-like relations for the kinematic factors associated to different
hypergeometric integrals \mafrabcj.

For example, one might start the CFT calculation using the OPE's of $U^2(z_2)$ 
to integrate out the conformal weight-one variables with $z_2$ dependence.
Then the OPE's of $U^3(z_3)$ and $U^4(z_4)$ can be used in different order
to get the $z_3$ and $z_4$ dependencies. The kinematic factors and integrands 
obtained with the two different orderings of OPE elimination
are simply related by relabeling $3 \leftrightarrow 4$ in their analogous expressions
for \manyLs.

While relabeling the kinematic factors is straightforward, relabeling the 
$z_{ij}$ dependencies may introduce different single-pole integrals which are not part 
of the original set of twenty-four obtained with the first ordering.
The end result of the CFT correlator being the same, there must be relations
which allow them to be expressed in terms of the original 
integrands\foot{The relations must hold for the $(z_{ij}z_{kl}z_{mn})^{-1}$ before doing the
$z_2,z_3,z_4$ integration 
because they refer to the properties of the CFT correlator. In particular, the total
derivative equations of Appendix B are not necessary to get BCJ-like relations for the
kinematic factors.}.
In fact, the partial fraction identities listed in (B.5) provide such
relations.
For example, the integrand $I_{13} \equiv 1/(z_{25}z_{34}z_{41})$ is relabeled to 
$1/(z_{25}z_{43}z_{31}) \equiv -I_{52}$, which is not in the original set. But
using (B.5) it can be rewritten as
\eqn\minusIcd{
I_{52} = I_{13} - I_{5},
}
and both $I_{13}$ and $I_5 \equiv 1/(z_{25}z_{31}z_{41})$ are present in \manyLs. 
By considering an augmented set of 63 integrands in Appendix B,
all new integrands
obtained via relabeling in this BCJ--inspired technique
can be rewritten in terms of linear combinations of
the original twenty-four\foot{This {\it augmented set
method} can be used to find the $(N-3)!$ basis of integrals for $N$--point
amplitude computations \ref\WIPS{C.R.~Mafra, S.~Stieberger, work in progress.}.}
integrands appearing in \manyLs.
Therefore, subtracting the amplitudes computed with different orderings for 
the OPE's gives relations among kinematic factors;
where some are originally present in \manyLs\ while others are simple relabellings 
of those. It turns out that this method allows the more involved
kinematic factors of \manyLs\ to be expressed in terms of simpler ones through
the following BCJ-like kinematic identities 
$$
 L_{213441} =  L_{214131} -  L_{213141}, \quad
 L_{213445} =  L_{214535} -  L_{213545}
$$
$$
 L_{233141} =  L_{312141} -  L_{213141}, \quad
 L_{233145} =  L_{312145} -  L_{213145}
$$
\eqn\BCJs{
 L_{233441} =  L_{213141} -  L_{312141} 
+  L_{413121} -  L_{412131}.
}
This method is particularly efficient in reducing the amount of
work. For example, the explicit computation of $L_{233441}$ is rather 
tedious because it involves OPE's
between three integrated vertices $U^i(z_i)$ among themselves,
as opposed to the simpler cases such as $L_{213141}$ and $L_{213545}$ 
where $U^i(z_i)$ collides with $V^j(z_j)$. 

Therefore the basis from which all 24 single-pole kinematic factors can
be obtained by simple relabelling is given by,
$$
\langle L_{213141}V_5 V_6\rangle,
$$
$$
\langle L_{213145} V_6\rangle =   \langle L_{54} L_{2131} V_6 \rangle,
$$
$$
\langle L_{213545} V_6\rangle =  \langle L_{3545}L_{12} V_6 \rangle,
$$
\eqn\genLsq{
\langle L_{213541} V_6\rangle =  \langle L_{53} L_{2141} V_6 \rangle,
}
where the equalities for the last three lines can be show by explicit
computations.

%**************************************************************
\subsec{Double pole integrands and total derivative techniques}
%**************************************************************
Similarly as in the 5-point computation, using the SYM equations of motion 
the kinematic factors of double-pole integrals can be rewritten
in such a way as to contain overall factors of $(1+ s_{ij})$. These are precisely
the factors which cancel the tachyon poles $(1+s_{ij})^{-1}$ in double-pole integrals and
allow them to be rewritten 
as linear combinations of single-pole integrals using total derivative 
relations of Appendix B. However, the six point amplitude additionally involves
integrals with tachyon poles $(1+s_{ijk})^{-1}$ in the $t_i$ variables which
are slightly more difficult to cancel.

It will be convenient to separate the double-pole contributions in \manyLs\ in
two distinct sets. Each of the last four terms of \manyLs\ 
\eqn\doubleD{\eqalign{
\int_0^1 dz_2 \int_{z_2}^1 dz_3 \int_{z_3}^{1} dz_4 &\prod_{i<j}^6 z_{ij}^{ - s_{ij}}\ \Big\{ 
{\langle L_{233434} V_1 V_5 V_6\rangle \over z_{23}   z_{34}^2}
+ {\langle L_{243434} V_1 V_5 V_6\rangle \over z_{24}   z_{34}^2}\cr
&+ {\langle L_{242434} V_1 V_5 V_6\rangle \over z_{24}^2 z_{34}}
+ {\langle L_{232334} V_1 V_5 V_6\rangle \over z_{23}^2 z_{34}}\Big\}}}
for itself is proportional
to the tachyon pole $(1+t_2)^{-1}$ due to the double-pole integrals 
$\cI_{27},\cI_{28},\cI_{29}$ and $\cI_{36}$. After some manipulations which are explained in appendix B.3, the double-pole contributions of \doubleD\ become
$$
\int_0^1 dz_2 \int_{z_2}^1 dz_3 \int_{z_3}^{1} dz_4 
\prod_{i<j}^6 z_{ij}^{ - s_{ij}}\ \Big\{
{\langle L_{233434} V_1 V_5 V_6\rangle \over z_{23}   z_{34}^2}
+ {\langle L_{243434} V_1 V_5 V_6\rangle \over z_{24}   z_{34}^2}
$$
\eqn\firstdoubles{
+ {\langle L_{242434} V_1 V_5 V_6\rangle \over z_{24}^2 z_{34}} + {\langle L_{232334} V_1 V_5 V_6\rangle \over z_{23}^2 z_{34}}\Big\} 
}
$$
= \langle  R_{243} V_1V_5V_6\rangle (- s_{12}\cI_{32} - s_{12}\cI_{16} + s_{12}\cI_{10} - s_{25}\cI_{33} - s_{25}\cI_{20} 
+ s_{25}\cI_{14} )
$$
$$
+ \langle R_{234}V_1V_5V_6\rangle (- s_{12}\cI_{30} + s_{12}\cI_{22} + s_{12}\cI_{10} - s_{12}\cI_{1} - s_{25}\cI_{31} 
+ s_{25}\cI_{26} + s_{25}\cI_{14} - s_{25}\cI_{8})
$$
$$
 + \langle \big[ D_{34}(k^3\cdot A^2) + D_{23}(k^2\cdot A^4) 
 - D_{24}(k^2\cdot A^3)\big]V_1V_5V_6\rangle (s_{14}\cI_{30}+s_{45}\cI_{31})
$$
$$
- \langle \big[ D_{34}(k^4\cdot A^2) - D_{23}(k^2\cdot A^4) + D_{24}(k^2\cdot A^3)\big]V_1V_5V_6\rangle
(s_{13}\cI_{32}+s_{35}\cI_{33}).
$$
These are simply corrections to the kinematics of the single-pole integrals with $R_{ijk}$ and $O_{ijk}$ building blocks.

The other set of double-pole integral contribution from \manyLs\ is given by
$$
\int_0^1 dz_2 \int_{z_2}^1 dz_3 \int_{z_3}^{1} dz_4 \prod_{i<j}^6 z_{ij}^{ - s_{ij}}\ \Big\{ 
{\langle L_{213434} V_5 V_6\rangle \over z_{21}   z_{34}^2}
+ {\langle L_{253434} V_1 V_6\rangle \over z_{25}   z_{34}^2}
+ {\langle L_{242431} V_5 V_6\rangle \over z_{24}^2 z_{31}}
$$
\eqn\otherDs{
+ {\langle L_{242435} V_1 V_6\rangle \over z_{24}^2 z_{35}}
+ {\langle L_{232341} V_5 V_6\rangle \over z_{23}^2 z_{41}}
+ {\langle L_{232345} V_1 V_6\rangle \over z_{23}^2 z_{45}}\ \Big\}.
}
By doing the explicit OPE computations with the conventions of section 2
one arrives at
$$
L_{213434} = (1+s_{34})L_{12}D_{34}, \quad L_{253434} = - (1+s_{34})L_{52}D_{34}, \quad
$$
$$
L_{232341} = (1+s_{23})L_{14}D_{23}, \quad L_{232345} = -(1+s_{23})L_{54}D_{23}, \quad
$$
\eqn\simpleDs{
L_{242431} = (1+s_{24})L_{13}D_{24}, \quad L_{242435} = -(1+s_{24})L_{53}D_{24}. \quad
}
The factors of $(1+s_{ij})$ in \simpleDs\ cancel all the tachyon poles 
and allow the application of total derivative relations from Appendix B
to rewrite \otherDs\ using only single-pole integrals.

%**********************************************************
\subsec{The six--point amplitude with BRST building blocks}
%**********************************************************

As explained in the previous subsection, the relations from Appendix B between 
generalized Euler integrals allow to rewrite the ten double pole integrals as corrections 
to the 24 single poles integrals. After going through a long computation one checks
that these double pole corrections are precisely of the form which replace the superfields
$L_{ij}, L_{jiki}$ and $L_{jikili}$ by their corresponding BRST building blocks $T_{ij}, T_{ijk}$
and $T_{ijkl}$ defined in subsection 2.2.
Furthermore, the resulting integrals accompanying these BRST kinematic factors 
can be rewritten using the partial fraction solutions (B.5),
e.g. $ \langle T_{1234}V_{5}V_{6} \rangle ({\cal I}_{30}
       - {\cal I}_{22}
       - {\cal I}_{10}
       + {\cal I}_1) = \langle T_{1234}V_{5}V_{6} \rangle \cI_{61}$,
leading to a surprinsingly simple answer,
\eqn\firstamp{
A(1,2,3,4,5,6) =
}
$$
  \langle T_{1234} V_5 V_6 \rangle \cI_{61}
- \langle T_{1243} V_5 V_6 \rangle \cI_{59} 
- \langle T_{1324} V_5 V_6 \rangle \cI_{55} 
- \langle T_{1342} V_5 V_6 \rangle \cI_{32}
$$
$$
- \langle T_{1423} V_5 V_6 \rangle \cI_{51}
+ \langle T_{1432} V_5 V_6 \rangle \cI_{30}
- \langle T_{5234} V_1 V_6 \rangle \cI_{43} 
+ \langle T_{5243} V_1 V_6 \rangle \cI_{40} 
$$
$$
+ \langle T_{5324} V_1 V_6 \rangle \cI_{45} 
+ \langle T_{5342} V_1 V_6 \rangle \cI_{33}       
+ \langle T_{5423} V_1 V_6 \rangle \cI_{44} 
- \langle T_{5432} V_1 V_6 \rangle \cI_{31}
$$
$$
- \langle T_{123}T_{45} V_6 \rangle \cI_{60}
- \langle T_{124}T_{35} V_6 \rangle \cI_{58}
+ \langle T_{132}T_{45} V_6 \rangle \cI_{25}
- \langle T_{134}T_{25}V_{6}\rangle \cI_{52}
$$
$$
        + \langle T_{142}T_{35}V_{6}\rangle {\cal I}_{19}
        + \langle T_{143}T_{25}V_{6}\rangle {\cal I}_{13}
        + \langle T_{253}T_{14}V_{6}\rangle {\cal I}_{49}
	+ \langle T_{254}T_{13}V_{6}\rangle {\cal I}_{53}
$$
$$
        - \langle T_{352}T_{14}V_{6}\rangle {\cal I}_{23}
        + \langle T_{354}T_{12}V_{6}\rangle {\cal I}_{57}
        - \langle T_{452}T_{13}V_{6}\rangle {\cal I}_{17}       
        - \langle T_{453}T_{12}V_{6}\rangle {\cal I}_{11}.
$$
Writing $\langle T_{ijkl} V_m V_n\rangle = -
\langle T_{ijkl} QT_{mn}\rangle /s_{mn}$ and integrating the
BRST charge by parts using \QTijkl, many terms cancel
due to the total derivative relations obeyed by the integrals and 
one arrives at the expression \simpleSpt\ presented in the Introduction,
\eqn\simpleSptd{
\eqalign{
A(1,2,3,4,5,6) &=
\langle T_{12}T_{34}T_{56}\rangle F'_1
+ \langle T_{54}T_{32}T_{16}\rangle F_1\cr
&+ \langle T_{14}T_{23}T_{56}\rangle F'_2
+ \langle T_{52}T_{43}T_{16}\rangle F_2
+ \langle T_{13}T_{24}T_{56}\rangle F'_3
+ \langle T_{53}T_{42}T_{16}\rangle F_3\cr
&+ \langle T_{123}E_{456}\rangle F_4
+ \langle T_{543}E_{216}\rangle F'_4
+ \langle T_{324}E_{561}\rangle F_5
+ \langle T_{342}E_{561}\rangle F'_5\cr
&+ \langle T_{435}E_{216}\rangle F_6
+ \langle T_{231}E_{456}\rangle F_6'
+ \langle T_{325}E_{416}\rangle F_7
+ \langle T_{341}E_{256}\rangle F'_7\cr
&+ \langle T_{124}E_{356}\rangle F_8
+ \langle T_{542}E_{316}\rangle F_8'
+ \langle T_{352}E_{416}\rangle F_9
+ \langle T_{314}E_{256}\rangle F_9'\cr
&+ \langle T_{241}E_{356}\rangle F_{10}
+ \langle T_{425}E_{316}\rangle F'_{10}\ ,
}}
where the integrals $F_i$ and $F'_i$ are defined by
\eqn\SSm{\eqalign{
F_1 &= -{ s_{13} {\cal I}_{25} + s_1 {\cal I}_{60}  \over s_6}, \quad
F_1'= -{ s_{35} {\cal I}_{57} + s_4 {\cal I}_{11} \over s_5}\ ,\cr
F_2 &= -{s_{14} {\cal I}_{13} + s_{13} {\cal I}_{52} \over s_6},\quad
F_2'= -{s_{25} {\cal I}_{49} + s_{35} {\cal I}_{23} \over s_5}\ ,\cr
F_3 &= -{s_{14} {\cal I}_{19} + s_1 {\cal I}_{58} \over s_6}, \quad 
F_3'= -{s_{25} {\cal I}_{53} + s_4 {\cal I}_{17}  \over s_5}\ ,\cr
F_4  &=  s_{4} {\cal I}_{2},\quad
F_4' = -s_{1} {\cal I}_{4},\quad
F_6 = s_{1} {\cal I}_{57},\quad
F_6'  = s_{4} {\cal I}_{25}\ ,\cr
F_7 &= s_{14} {\cal I}_{49},\quad
F_7' = s_{25} {\cal I}_{13},\quad
F_8 = s_{35} {\cal I}_{3},\quad
F_8' =-s_{13} {\cal I}_6\ ,\cr
F_9 &= s_{14} {\cal I}_7, \quad
F_9' = -s_{25} {\cal I}_5 , \quad
F_{10} = s_{35} {\cal I}_{19} , \quad
F_{10}' = s_{13} {\cal I}_{53}\ ,\cr
F_5&= s_1 {\cal I}_{62} + s_{13} {\cal I}_{55} + s_{14} {\cal I}_{51}, \quad
F_5' = s_1 {\cal I}_{59} + s_{13} {\cal I}_{32} + s_{14} {\cal I}_{48}.
}}
The explicit $\a'$--expansions of the integrals $F_i$ obtained using the methods explained
in \StieOpr\ are written down in Appendix B.

%**************************
\subsec{Field theory limit}
%**************************

Plugging in the momenta expansions for the $F_i, F'_i$ integrals appearing in \simpleSpt\ and
taking their limit as $\a' \rightarrow 0$ gives the field theory super-Yang-Mills six--point tree 
amplitude
\eqn\FTlimit{
A_{\rm SYM}(1,2,3,4,5,6) =
 {\langle T_{12}T_{34}T_{56}\rangle \over s_1 s_3 s_5}
+ {\langle T_{54}T_{32}T_{16}\rangle\over s_2 s_4 s_6}
}
$$
+ {\langle T_{123}E_{456}\rangle\over s_1 t_1}
+ {\langle T_{543}E_{216}\rangle\over s_4 t_3}
+ {\langle T_{321}E_{456}\rangle \over s_2 t_1}
+ {\langle T_{345}E_{216}\rangle \over s_3 t_3}
+ {\langle T_{432}E_{561}\rangle \over s_3 t_2}
+ {\langle T_{234}E_{561}\rangle \over s_2 t_2},
$$
which agrees with the Ansatz proposed in \FTAmps, and therefore proves it.
To see this first rewrite the six--point expression of \FTAmps\ using the 
$E_{ijk}$ building blocks to obtain
\eqn\sAnsatz{
A_{\rm SYM}^{\rm ansatz}(1,2,3,4,5,6) =  {\langle T_{12}T_{34}T_{56}\rangle \over s_1 s_3 s_5}
+ {\langle T_{54}T_{32}T_{16}\rangle\over s_2 s_4 s_6}
}
$$
+\half\big( {T_{123}E_{456}\over s_1 t_1} + {T_{234}E_{561}\over s_2 t_2} + {T_{345}E_{612}\over s_3 t_3}
+ {T_{456}E_{123}\over s_4 t_1}  + {T_{561}E_{234}\over s_5 t_2} + {T_{612}E_{345}\over s_6 t_3}\big)
$$
$$
-\half\big( {T_{126}E_{345}\over s_1 t_3} + {T_{231}E_{456}\over s_2 t_1} + {T_{342}E_{561}\over s_3 t_2}
+ {T_{453}E_{612}\over s_4 t_3}  + {T_{564}E_{123}\over s_5 t_1} + {T_{615}E_{234}\over s_6 t_2}\big).
$$
To relate the last two lines of \sAnsatz\ with the last line of \FTlimit\ it is sufficient
to consider the identities
\eqn\brstvan{
0 = \langle Q\big[ \({T_{123}\over s_1} + {T_{321}\over s_2}\)
\({T_{456}\over s_4} + {T_{654}\over s_5}\)\big]\rangle, \quad
Q\big[ {T_{123}\over s_{12}} + {T_{321}\over s_{23}}\big] = - s_{123}E_{123},
}
and relabellings of it. Using \brstvan\ and the momentum conservation relation 
$s_{123} = s_{456}$ one obtains, for example
\eqn\prima{
\langle \({T_{456}\over s_4} - {T_{564}\over s_5}\) E_{123}\rangle
=  \langle \({T_{123}\over s_1} - {T_{231}\over s_2}\) E_{456}\rangle,
}
which together with its relabellings allow one to prove that \sAnsatz\ and \FTlimit\
are equal;
$A_{\rm SYM}^{\rm ansatz} = A_{\rm SYM}$. Each term of the field theory
limit expression \FTlimit\ can be associated to Feynman diagrams which use only
three--point vertices as in the arguments of \BCJ. The explicit mapping is shown in
Figure 4 of the Appendix D.

%************************************
\newsec{BCJ identities in superspace}
%************************************

In field theory 
the kinematic factors of an $N$--point tree--level gluon amplitude can be 
rearranged such that the form of any partial amplitude becomes rather simple. 
More precisely, kinematic factors corresponding to diagrams with 
purely cubic interactions can be chosen and any subamplitude is 
organized as a sum over terms describing these diagrams~\BCJ.
The latter are specified by some numerator factors and their corresponding propagator structure. 
The total number of such numerator factors or channels
is given by $(2N-5)!!$, while the number of independent factors is $(N-2)!$ \BCJ.
Any contact term may be absorbed into the numerator factors of the diagrams.
Moreover, it has been argued in~\BCJ, that  kinematic numerator identities 
impose additional constraints. As a consequence in field--theory the number of independent color--ordered
$N$--point amplitudes at tree--level is $(N-3)!$ \BCJ.
As demonstrated in \refs{\BjerrumBohrRD,\StiebergerHQ} this result may be easily anticipated from string theory.

%********************************************************
\subsec{Extended BCJ relations from monodromy  relations}
%********************************************************
\def\h{{1\over 2}}
\def\si{\sigma}

After imposing cyclic symmetry, reflection and parity symmetries
there are $\h(N-1)!$ different color ordered subamplitudes.
Furthermore, after applying Kleiss--Kuijf relations we end up at $(N-2)!$ independent
subamplitudes.
As a basis we may choose the  following $(N-2)!$ elements
$A(1,2_\si,{\ldots},(N-1)_\si,N)$ with the permutation
$\si \in S_{N-2}$ \refs{\KK,\KKi}.
Hence, for the case of interest with $N=6$ we need to specify
$24$ subamplitudes and parametrize the latter according to their 
pole structure, which arises from diagrams with only three--vertices contributing.
To each amplitude $14$ terms or diagrams contribute. The topological 
structure of the latter is depicted in the following two figures.
\ifig\figii{Diagrams giving rise to 
the numerator factor $n_i$.}
{\epsfxsize=1\hsize\epsfbox{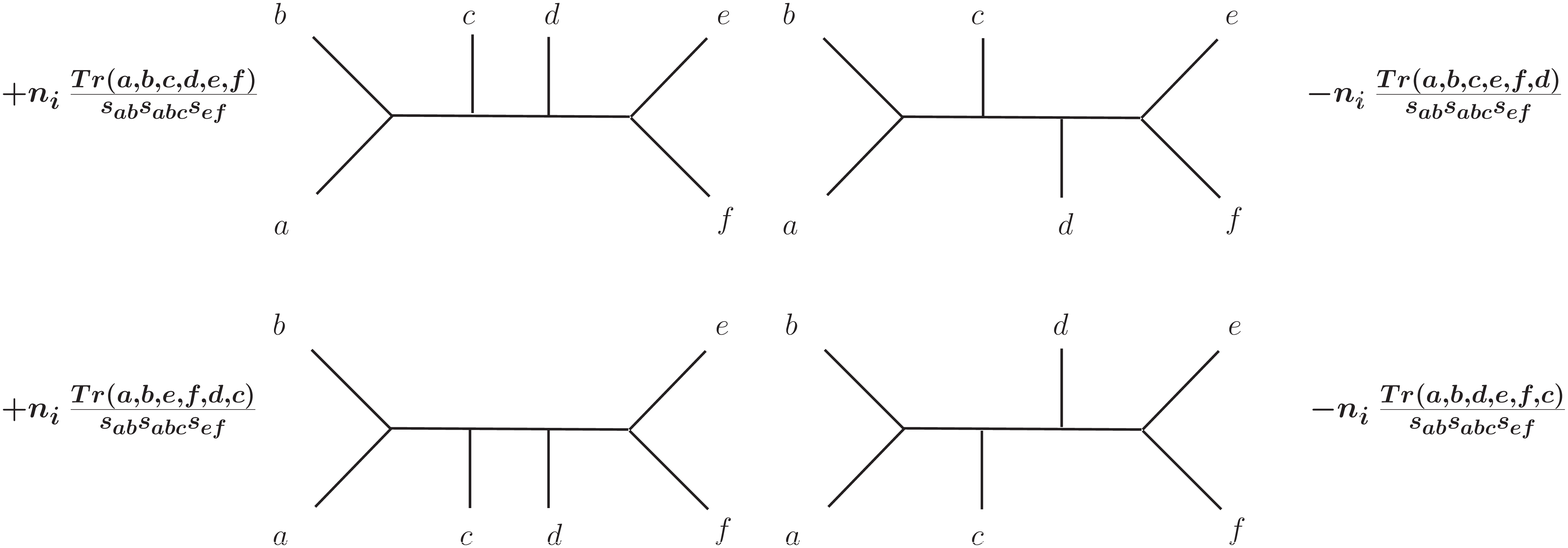}}
\ifig\figiii{Diagrams giving rise to 
the numerator factor $n_j$.}
{\epsfxsize=1\hsize\epsfbox{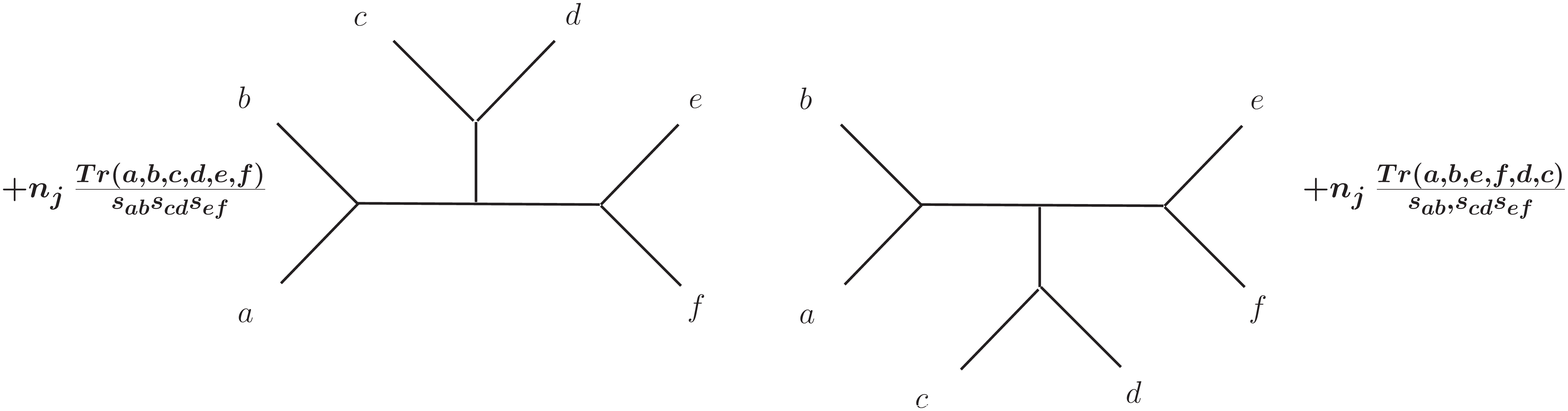}}
\vskip0.5cm
\noindent For the first amplitude we make the Ansatz
\def\crr{\noalign{\vskip5pt}}
\eqn\Ansatz{\eqalign{
A(1,2,3,4,5,6)&= 
 {n_{1}(2345)\over s_{12} s_{34} s_{56}}
+{n_{2}(2345)\over  s_{23} s_{45} s_{123}}
+{n_{3}(2345)\over  s_{12} s_{45} s_{123}}
+{n_{4}(2345)\over  s_{12} s_{56} s_{123}}
+{n_{5}(2345)\over s_{23} s_{56} s_{123}}\cr\crr
&+{n_{6}(2345)\over s_{23} s_{56} s_{156}}
+{n_{7}(2345)\over s_{34} s_{56} s_{156}}
+{n'_1(2345)\over s_{16} s_{23} s_{45}}
+{n'_2(2345)\over s_{12} s_{34} s_{126}}
+{n'_3(2345)\over s_{12} s_{45} s_{126}}\cr\crr
&+{n'_4(2345)\over s_{16} s_{45} s_{126}}
+{n'_5(2345)\over s_{16} s_{34} s_{126}}
+{n'_6(2345)\over s_{16} s_{34} s_{156}}
+{n'_7(2345)\over s_{16} s_{23} s_{156}}\ ,}}
with the numerator factors $n_i$.
In the remaining $23$ subamplitudes we have to take into account, that 
a numerator factor $n_i$ may contribute to various other subamplitudes.
In Fig. 2 and Fig. 3 we display diagrams, 
which give rise to the same numerator factors $n_i$ and $n_j$, respectively. 
Taking this fact into account yields:
\eqn\Ansatzi{\eqalign{
A(1,2,3,5,4,6)&=
  {n_{1}(2354)\over s_{12} s_{35} s_{46}}
- {n_{2}(2345)\over s_{23} s_{45} s_{123}}
- {n_{3}(2345)\over s_{12} s_{45} s_{123}}
+ {n_{4}(2354)\over s_{12} s_{46} s_{123}}
+ {n_{5}(2354)\over s_{23} s_{46} s_{123}}\cr\crr
&+ {n_{6}(2354)\over s_{23} s_{46} s_{146}}
+ {n_{7}(2354)\over s_{35} s_{46} s_{146}}
- {n'_1(2345)\over s_{16} s_{23} s_{45}}
+ {n'_2(2354)\over s_{12} s_{35} s_{126}}
- {n'_3(2345)\over s_{12} s_{45} s_{126}}\cr\crr
&- {n'_4(2345)\over s_{16} s_{45} s_{126}}
+ {n'_5(2354)\over s_{16} s_{35} s_{126}}
+ {n'_6(2354)\over s_{16} s_{35} s_{146}}
+ {n'_7(2354)\over s_{16} s_{23} s_{146}}\ ,\cr
\vdots\cr
A(1,5,4,3,2,6)&=
- {n_{1}(5342)\over s_{15} s_{26} s_{34}}
+ {n_{2}(4523)\over s_{23} s_{45} s_{145}}
- {n_{3}(5423)\over s_{15} s_{23} s_{145}}
+ {n_{4}(5432)\over s_{15} s_{26} s_{145}}
- {n_{5}(4532)\over s_{26} s_{45} s_{145}}\cr\crr
&+ {n_{6}(3452)\over s_{26} s_{34} s_{126}}
+ {n_{7}(3452)\over s_{26} s_{45} s_{126}}
- {n'_1(2345)\over s_{16} s_{23} s_{45}}
+ {n'_2(5234)\over s_{15} s_{23} s_{156}}
+ {n'_3(5234)\over s_{15} s_{34} s_{156}}\cr\crr
&- {n'_4(2345)\over s_{16} s_{45} s_{126}}
- {n'_5(2345)\over s_{16} s_{34} s_{126}}
- {n'_6(2345)\over s_{16} s_{34} s_{156}}
- {n'_7(2345)\over s_{16} s_{23} s_{156}}\ .}}
The parametrization of the remaining $21$ subamplitudes is given in Appendix C.
In total we need $7!!=105$ numerator factors $n_i$ to parametrize the $24$
subamplitudes.

Now we shall make use of the monodromy  relations, which give rise to 
non--trivial relations between various different subamplitudes 
\refs{\BjerrumBohrRD,\StiebergerHQ}.
One of these relations reads \StiebergerHQ:
\eqn\Monodr{\eqalign{
A(1,2,3,&4,5,6)+e^{i\pi s_{12}}\ A(2,1,3,4,5,6)+e^{i\pi(s_{12}+s_{13})}\ 
A(2,3,1,4,5,6)\cr
&+e^{i\pi(s_{12}+s_{13}+s_{14})}\ A(2,3,4,1,5,6)+
e^{i\pi(s_{12}+s_{13}+s_{14}+s_{15})}\ A(2,3,4,5,1,6)=0\ .}}
A complete set can be obtained by permuting all open string labels.
In the field--theory limit the real part of all these relations gives rise 
to the Kleiss--Kuijf relations.
Hence, these relations allow to determine all $60$ partial subamplitudes
from the set of $24$ given in \Ansatz\ and \Ansatzi.
On the other hand, it has been argued in \refs{\tye,\BjerrumBohrZS}, 
that the imaginary part of all relations gives rise to a set of equations, 
the so--called extended BCJ relations, which relate the numerator factors. 
In these equations three numerator factors $n_i$ constitute the triplets $X_j$. 
In the following we define the hundred triplets:
\eqn\NNs{\eqalign{
X_{1}&=n_{1}(2 3 4 5)-n_{4}(2 3 4 5)+n_{4}(2 4 3 5)\ \ \ ,\ \ \  
X_{2}=n_{2}(2 3 4 5)-n_{5}(2 3 4 5)+n_{5}(2 3 5 4) \ ,\cr
&\vdots\cr
X_{99}&=n_4(4532)-n_4(5432)-n_5(4532)\ \ \ ,\ \ \ 
X_{100}=-n_1(5342)+n_5(3452)-n_6(3452)\ .}}
The remaining $96$ triplets can be found in Appendix C.
{}From the imaginary part of the monodromy  relations an independent set 
of $3!\cdot 3=18$ equations each containing $15$ triplets \NNs\ can be derived:
\eqn\extBCJs{\eqalign{
{X_{1}\over s_{1 2} s_{5 6}}&
-{X_{2}\over s_{2 3} s_{1 2 3}}
-{X_{3}\over s_{1 2} s_{1 2 3}}
-{X_{4}\over s_{2 3} s_{5 6}}
+{X_{5}\over s_{5 6} s_{1 5 6}}
-{X_{6}\over s_{2 3} s_{6 1}}
+{X_{7}\over s_{1 2} s_{1 2 6}}
-{X_{8}\over s_{6 1} s_{1 2 6}}\cr\crr
&+{X_{9}\over s_{6 1} s_{1 5 6}}
+{X_{10}\over s_{1 2} s_{3 5}}
+{X_{11}\over s_{3 5} s_{6 1}}
+{X_{12}\over s_{3 5} s_{1 2 4}}
+{X_{13}\over s_{5 6} s_{1 2 4}}
+{X_{14}\over s_{3 5} s_{1 4 6}}
-{X_{15}\over s_{2 3} s_{1 4 6}}=0\ ,\cr
&\vdots\cr
{X_{4}\over s_{2 3} s_{5 6}}&
+{X_{20}\over s_{2 3} s_{1 4 5}}
-{X_{26}\over s_{4 5} s_{1 2 3}}
-{X_{27}\over s_{5 6} s_{1 2 3}}
-{X_{44}\over s_{4 5} s_{1 2 6}}
-{X_{45}\over s_{3 4} s_{1 2 6}}
+{X_{52}\over s_{3 4} s_{5 6}}
-{X_{64}\over s_{2 3} s_{4 5}}\cr\crr
&-{X_{70}\over s_{2 6} s_{4 5}}
+{X_{89}\over s_{2 6} s_{1 3 4}}
-{X_{82}\over s_{5 6} s_{1 3 4}}
+{X_{93}\over s_{2 3} s_{1 5 6}}
+{X_{98}\over s_{3 4} s_{1 5 6}}
+{X_{99}\over s_{2 6} s_{1 4 5}}
-{X_{100}\over s_{2 6} s_{3 4}}=0\ .}}
The remaining $16$ equations are listed in Appendix C.
In the relations \extBCJs\ each triplet $X_i$ is the numerator 
of a product of $N-4=2$ poles $s_I, s_J$.
Eqs. \extBCJs\ imply that the residue of each pole term must vanish, i.e.:
\eqn\residue{
\left.X_i\ \right|_{\rm residue}=0\ .}
However, the regular part of the triplets $X_i$, which is proportional to $s_I s_J$, may be non--vanishing.
The choice $X_i=0$ would also be a solution of 
the equations \extBCJs, but it corresponds to a specific gauge choice.
%%  Their solution is explicitly displayed in Appendix D.
The  equations \residue\ correspond to the set of color identities \BCJ.
In fact, these identities give rise to $81$ independent kinematic equations 
relating the $105$ numerators $n_i$. Hence, in total there are $24$ independent 
numerators $n_i$.
On the other hand, the set of $18$ extended BCJ relations \extBCJs\ describes the
general constraint on the numerator factors. 

As will be demonstrated below, using the field--theory parametrization 
following from \firstamp\
together with the hook properties of the pure spinor building blocks 
$T_{ijk}$ and $T_{ijkl}$
it is possible to easily identify the explicit
$(N-2)!$ basis numerators. The explicit form of the supersymmetric 
numerator factors $n_i$ in \Ansatz\ can be read off by comparing it with \FTlimit.

%**********************************
\subsec{Basis numerators for $N=5$}
%**********************************

The tree--level amplitude prescription \sixpt\ from string theory naturally suggests
using the $(N-3)!$ subamplitudes generated by the different orderings of the integrated
vertices as a basis, i.e. $A(1,2_\s,{\ldots},(N-2)_\s,N-1,N)$ with $\s \in S_{N-3}$ and positions $(1,N-1,N)$ fixed.
For example, using the field--theory limit of the five--point amplitudes computed
with the pure spinor formalism in \refs{\FTAmps,\mafrabcj}, the $(N-3)!=2$ basis amplitudes
can be written in terms of ten kinematic factors as\foot{With
the notation of \BjerrumBohrZS\ one can show that this parametrization leads
to the vanishing of four triplets: $X_3=X_5 = X_7 = n_{14} - n_{12} + n_{13} =0$ and the extended BCJ identities are satisfied.}:
$$
A_{\rm SYM}(1,2,3,4,5) =  
            {\langle T_{123} V^4 V^5\rangle  \over s_{12} s_{45}}
          - {\langle T_{234} V^1 V^5\rangle \over s_{23} s_{51}}
	  + {\langle L_{12} L_{34} V^5\rangle \over s_{12} s_{34}}
          + {\langle T_{321} V^4 V^5\rangle \over s_{23} s_{45}}
          - {\langle T_{432} V^1 V^5\rangle \over s_{34} s_{51}},
$$
\eqn\fivetwo{
A_{\rm SYM}(1,3,2,4,5) =  
            {\langle T_{132} V^4 V^5\rangle  \over s_{13} s_{45}}
          - {\langle T_{324} V^1 V^5\rangle \over s_{23} s_{51}}
	  + {\langle L_{13} L_{24} V^5\rangle \over s_{13} s_{24}}
          + {\langle T_{231} V^4 V^5\rangle \over s_{23} s_{45}}
          - {\langle T_{423} V^1 V^5\rangle \over s_{24} s_{51}}.
}
However, the number of independent kinematic factors is $(N-2)!=6$ because
of the hook symmetries \Tijkhooks\ of $T_{ijk}$,
$$
\langle T_{324} V^1 V^5\rangle = - \langle T_{234} V^1 V^5\rangle, \quad
\langle T_{231} V^4 V^5\rangle = - \langle T_{321} V^4 V^5\rangle,
$$
$$
\langle T_{132} V^4 V^5\rangle = \langle T_{123} V^4 V^5\rangle - \langle T_{321} V^4 V^5\rangle,
$$
\eqn\hooks{
\langle T_{423} V^1 V^5\rangle =  \langle T_{432} V^1 V^5\rangle - \langle T_{234} V^1 V^5\rangle.
}

%**********************************
\subsec{Basis numerators for $N=6$}
%**********************************

The six--point subamplitude in the field-theory limit \sAnsatz\
is expanded in terms of 14 poles, so the full $(N-3)!=6$ basis amplitudes 
would naively correspond to 84 kinematic factors. However,
the pure spinor BRST building block form of the kinematic factors 
allows one to easily find the basis with $(N-2)!=24$ elements, in accord with
the monodromy analysis of section 4.1.

To see this
it is convenient to use the field-theory limit of \firstamp, 
\eqn\sixptBCJ{
A_{\rm SYM}(1,2,3,4,5,6) =
   {\langle (T_{1234} - T_{1243}) V_5 V_6\rangle \over s_1 s_3 s_5}
 + {\langle (T_{5432} - T_{5423}) V_1 V_6\rangle \over s_2 s_4 s_6}
}
$$
 + {\langle (T_{132} - T_{123}) T_{45} V_6\rangle \over s_2 s_4 t_1}
 + {\langle (T_{534} - T_{543}) T_{21} V_6\rangle \over  s_1 s_3 t_3}
 - {\langle T_{123} L_{45} V_6\rangle \over s_1 s_4 t_1}           
 - {\langle T_{453} L_{12} V_6\rangle \over s_1 s_4  t_3}
$$
$$
       + {\langle T_{1234} V_5V_6\rangle \over  s_1 s_5 t_1}
       + {\langle T_{5432} V_1V_6\rangle \over s_4s_6t_3}
       + {\langle (T_{1234} - T_{1324}) V_5V_6\rangle \over s_2s_5t_1}
       + {\langle (T_{5432} - T_{5342}) V_1V_6\rangle \over s_3s_6t_3}
$$
$$
        + {\langle (T_{1234} - T_{1324} - T_{1423} + T_{1432}) V_5V_6\rangle \over s_2s_5t_2}
        + {\langle (T_{5432} - T_{5342} - T_{5243} + T_{5234}) V_1V_6\rangle \over  s_3s_6t_2}
$$
$$
        + {\langle (T_{1234} - T_{1243} - T_{1342} + T_{1432}) V_5V_6\rangle \over  s_3s_5t_2}
	+ {\langle (T_{5432} - T_{5423} - T_{5324} + T_{5234}) V_1V_6\rangle \over s_2s_6t_2}.
$$
The field-theory amplitude \sixptBCJ\ is more conveniently written as
$$
A(1,2,3,4,5,6) =
   {n_1(234)  \over s_1 s_3 s_5}
 + {n'_1(234) \over s_2 s_4 s_6}
 + {n_2(234)  \over s_2 s_4 t_1}
 + {n'_2(234) \over s_1 s_3 t_3}
 + {n_3(234)  \over s_1 s_4 t_1}           
 + {n'_3(234) \over s_1 s_4 t_3}
$$
\eqn\fourteen{
 + {n_4(234)  \over  s_1s_5t_1}
 + {n'_4(234) \over  s_4s_6t_3}
 + {n_5(234)  \over  s_2s_5t_1}
 + {n'_5(234) \over  s_3s_6t_3}
 + {n_6(234)  \over  s_2s_5t_2}
 + {n'_6(234) \over  s_3s_6t_2}
 + {n_7(234)  \over  s_3s_5t_2}
 + {n'_7(234) \over  s_2s_6t_2},
}
where $n_i(jkl)\equiv n_i(jkl5)$ such that the labels $(jkl)$ denote the ordering of the 
integrated vertices in the $(N-3)!=6$ basis of
partial amplitudes. Their explicit form in terms of pure spinor BRST building blocks read
$$
n_1(234) = \langle(T_{1234} - T_{1243}) V_5 V_6 \rangle,\quad
n_2(234) = \langle (T_{132} - T_{123}) T_{45} V_6\rangle,
$$
$$
n_3(234) =  - \langle T_{123} T_{45} V_6\rangle,\quad
n_4(234) =    \langle T_{1234} V_5V_6\rangle,\quad
n_5(234) = \langle (T_{1234} - T_{1324}) V_5V_6\rangle,
$$
$$
n_6(234) = \langle (T_{1234} - T_{1324} - T_{1423} + T_{1432}) V_5V_6\rangle,
$$
\eqn\ndefs{
n_7(234) = \langle (T_{1234} - T_{1243} - T_{1342} + T_{1432}) V_5V_6\rangle,
}
while $n'_i(ijk)$ is obtained from $n_i(ijk)$ by the parity transformation $1\leftrightarrow 5, 2\leftrightarrow 4$.

It is now straightforward to check that the set of $12\oplus 12'$ kinematic factors
\eqn\setkin{
\{ n_3(ijk),\, n_4(ijk),\, n'_3(ijk),\, n'_4(ijk)\; /\, (ijk) \in {\rm perm}(234) \}
}
is a basis from which all 84 kinematic numerators can be obtained. In fact, the explicit
BCJ-like solution reads
$$
n_1(234) = n_4(234) - n_4(243),\quad n_2(234) = n_3(234) - n_3(243),\quad
$$
$$
n_5(234) = n_4(234) - n_4(324),\quad n_6(234) = n_4(234) - n_4(324) - n_4(423) + n_4(432)
$$
\eqn\solBCJ{
n_7(234) =  n_4(234) - n_4(243) - n_4(342) + n_4(432).
}
Together with the permutations of $(234)$ (with corresponding equations for $n'_i(ijk)$),
the solution \solBCJ\ generates $30\oplus 30'$ equations which allow reducing 
the 84 kinematic factors down to 24. One can also show that using the parametrization
given by \sixptBCJ\ together with the hook properties of $T_{ijk}$ and $T_{ijkl}$, 
59 triplets defined in \NNs\ trivially vanish
\eqn\vanishX{
X_i = 0, \quad i=1,4-9,11-13,16-18,20,21,24,26-28,30,31,35-39,
}
$$
42,52-56,62,63,70-75,77,78,80-82,84-92,95-97,99,100.
$$
By applying \PSS\ the 18 extended BCJ's \extBCJs\ have  explicitly been checked  
to be satisfied using the pure spinor representation \ndefs\  for the amplitudes in \Ansatzi. In the Appendix~C it
is explicitly shown that using the symmetry properties of $T_{ijk}$ and $T_{ijkl}$,
the solution to $X_i=0$ for $i=1,{\ldots},100$ implies that all $105$ numerators $n_j$ in
the Kleiss-Kuijf basis can be expressed in terms of the basis \setkin.

\goodbreak
\vskip 5mm
\centerline{\noindent{\bf Acknowledgments} }

We are grateful to Giuseppe Policastro for useful discussions.
C.M. would like to thank the Werner--Heisenberg--Institut 
in M\"unchen for hospitality and the friendly atmosphere during preparation of this work. 
C.M. and D.T. also thank the organizers of the
Amsterdam String Theory Workshop, where some discussions took place in
an inspiring location.
C.M. thanks the partial financial support from
the MPG and acknowledges support by the Deutsch-Israelische
Projektkooperation (DIP H52).
St.St. would like to thank  the Albert--Einstein--Institut in Potsdam and in particular 
Hermann Nicolai and Stefan Theisen for invitation and partial support during completion of this work.
The diagrams have been created by the program JaxoDraw~\Jaxo.

%\break
%************************************************************************
\appendix{A}{The superfields $L_{21}$, $L_{2131}$ and $L_{213141}$}
%************************************************************************

Using the OPE's of section 2 and a few BRST manipulations together with the
SYM equations of motion it follows that\foot{It was pointed out in \refs{\PSSids,\mafrabcj} that
extracting $L_{ij}$ and $L_{jiki}$ from the $V^iU^j$ and $V^iU^jU^k$ OPE's additionally involve
$Q$ integration by parts. The same happens for $L_{jikili}$, but the
arising extra terms of schematic form $s_{ij}(A^k W^l)$, $s_{ij}(A^k \g^{pq}W^l)$ and
$s_{ij}(W^k {\slashchar A_l} W^m)$ 
turn out to cancel in the overall 6pt amplitude.}
\eqn\Lot{
\lim_{z_2 \to z_1} V^1(z_1)U^2(z_2) \rightarrow {L_{12}\over z_{12}}, \quad
\lim_{z_2,z_3 \to z_1} V^1(z_1)U^2(z_2)U^3(z_3) \rightarrow {L_{2131}\over z_{21}z_{31}},
}
where \refs{\PSSids, \mafrabcj}
$$
L_{12} = A^1_m (\l\g^m W^2) + V^1(k^1\cdot A^2),
$$
\eqn\Ltoto{
L_{2131} = L_{12}((k^1+k^2)\cdot A^3)
+  (\l\g^{m}W^3) \big[  A^1_m(k^1\cdot A^2) +  A^{1\,n}{\cal F}^2_{mn}
       - (W^1\g_m W^2)\big]
}
Finally, a long computation leads to
\eqn\Ltotofodef{
\lim_{z_2,z_3,z_4 \to z_1} V^1(z_1)U^2(z_2)U^3(z_3)U^4(z_4) \rightarrow {L_{213141}\over z_{21}z_{31}z_{41}}
}
where 
\eqn\Ltotofo{
L_{213141} = - L_{2131}  \big[  A_4 \cdot (k_1+k_2+k_3)  \big] +  (A_3 \cdot (k_1 + k_2))
}
$$
  \times \big[(A_1 \cdot k_2)  A_m^2  
 -  (A_1 \cdot A_2)  k_m^2
 -  A_m^1 (A_2 \cdot k_1)  
 +  (W_1 \g_m W_2) \big](\l \g^m W_4)
$$
$$
 + (\l \g^m W_4) \big[ (A_1 \cdot k_2)  (A_2 \cdot A_3)  k_m^3 
 -  (A_1 \cdot k_2)  (W_2 \g^m W_3) 
 -  (A_1 \cdot A_2)  (k_2 \cdot A_3)  k_m^3
$$
$$
 +  (A_1 \cdot A_2)  (k_2 \cdot k_3)  A_m^3  
 -  (A_1 \cdot k_2)  (A_2 \cdot k_3)  A_m^3
 +  (A_1 \cdot k_3)  (k_1 \cdot A_2)  A_m^3 
 -  (A_1 \cdot A_3)  (k_1 \cdot A_2)  k_m^3 
$$
$$
 +  (W_1 \g^n W_2)  {\cal F}_{mn}^3
 +  {1\over 4}  (W_2 \g_{pq} \g_m W_3)  {\cal F}_1^{pq}
 -  {1\over 4}  (W_1 \g_{pq} \g_m W_3)  {\cal F}_2^{pq}
 +  (W_1 \g_m W_3) (k_1 \cdot A_2) \big].
$$

%*********************************
\appendix{B}{Six--point integrals}
%*********************************

A direct computation of the six--point amplitude with the pure spinor formalism
requires 34 triple integrals of the form
\eqn\intdef{
{\cal I}_k = \int dz_2 \int dz_3 \int dz_4 \prod_{i<j}^6 z_{ij}^{ - s_{ij}} I_k,
}
where $I_k,\; k = 1,{\ldots},34$ will be written below.
However, it is convenient to consider an augmented set of 63 integrals by including
$I_k,\; k=35,{\ldots}, 63$ which allow the definition of a system of equations which
can be used to simplify the amplitude considerably. This convenient set of $\{I_k\}$
is given by
$$
I_1 = {1\over z_{21} z_{31} z_{41} }, \quad 
I_2 = {1\over z_{21} z_{31} z_{45} }, \quad 
I_3 = {1\over z_{21} z_{35} z_{41} }, \quad 
I_4 = {1\over z_{21} z_{35} z_{45} }
$$
$$
I_5 = {1\over z_{25} z_{31} z_{41} }, \quad 
I_6 = {1\over z_{25} z_{31} z_{45} }, \quad
I_7 = {1\over z_{25} z_{35} z_{41} }, \quad 
I_8 = {1\over z_{25} z_{35} z_{45} } 
$$
$$
I_9 = {1\over z_{21} z_{34}^2 }, \quad\quad
I_{10} = {1\over z_{21} z_{34} z_{41} }, \quad
I_{11} = {1\over z_{21} z_{34} z_{45} }, \quad 
I_{12} = {1\over z_{25} z_{34}^2 }
$$
$$
I_{13} = {1\over z_{25} z_{34} z_{41} }, \quad 
I_{14} = {1\over z_{25} z_{34} z_{45} }, \quad
I_{15} = {1\over z_{31} z_{24}^2 }, \quad 
I_{16} = {1\over z_{31} z_{24} z_{41} } 
$$
$$
I_{17} = {1\over z_{31} z_{24} z_{45} }, \quad
I_{18} = {1\over z_{35} z_{24}^2 }, \quad 
I_{19} = {1\over z_{35} z_{24} z_{41} }, \quad 
I_{20} = {1\over z_{35} z_{24} z_{45} }
$$
$$
I_{21} = {1\over z_{41} z_{23}^2 }, \quad\quad
I_{22} = {1\over z_{41} z_{23} z_{31} }, \quad 
I_{23} = {1\over z_{41} z_{23} z_{35} }, \quad
I_{24} = {1\over z_{45} z_{23}^2 }
$$
$$
I_{25} = {1\over z_{45} z_{23} z_{31} }, \quad 
I_{26} = {1\over z_{45} z_{23} z_{35} }, \quad
I_{27} = {1\over z_{23} z_{34}^2 }, \quad\quad 
I_{28} = {1\over z_{24} z_{34}^2 } 
$$
$$
I_{29} = {1\over z_{34} z_{24}^2 }, \quad\quad
I_{30} = {1\over z_{23} z_{34} z_{41} }, \quad 
I_{31} = {1\over z_{23} z_{34} z_{45} }, \quad 
I_{32} = {1\over z_{24} z_{34} z_{31} }
$$
$$
I_{33} = {1\over z_{24} z_{34} z_{35} }, \quad
I_{34} = {1\over  z_{23}^2 z_{24} }, \quad\quad 
I_{35} = {1\over z_{24}^2 z_{23} }, \quad\quad
I_{36} = {1\over z_{23}^2 z_{34} } \quad
$$
$$
I_{37}        = {1\over z_{25} z_{34} z_{35} }, \quad 
I_{38}        = {1\over z_{24} z_{34} z_{45} }, \quad 
I_{39}        = {1\over z_{24} z_{25} z_{35} }, \quad 
I_{40}        = {1\over z_{24} z_{25} z_{34} } 
$$
$$
I_{41}        = {1\over z_{23} z_{34} z_{35} }, \quad 
I_{42}        = {1\over z_{23} z_{25} z_{45} }, \quad 
I_{43}        = {1\over z_{23} z_{25} z_{34} }, \quad 
I_{44}        = {1\over z_{23} z_{24} z_{45} } 
$$
$$
I_{45}        = {1\over z_{23} z_{24} z_{35} }, \quad 
I_{46}        = {1\over z_{23} z_{24} z_{34} }, \quad 
I_{47}        = {1\over z_{23} z_{24} z_{25} }, \quad 
I_{48}        = {1\over z_{41} z_{24} z_{34} } 
$$
$$
I_{49}        = {1\over z_{41} z_{23} z_{25} }, \quad 
I_{50}        = {1\over z_{21} z_{31} z_{24} }, \quad 
I_{51}        = {1\over z_{41} z_{23} z_{24} }, \quad 
I_{52}        = {1\over z_{31} z_{25} z_{34} } 
$$
$$
I_{53}        = {1\over z_{31} z_{24} z_{25} }, \quad 
I_{54}        = {1\over z_{31} z_{23} z_{34} }, \quad 
I_{55}        = {1\over z_{31} z_{23} z_{24} }, \quad 
I_{56}        = {1\over z_{21} z_{31} z_{34} } 
$$
$$
I_{57}        = {1\over z_{21} z_{34} z_{35} }, \quad 
I_{58}        = {1\over z_{21} z_{24} z_{35} }, \quad 
I_{59}        = {1\over z_{21} z_{24} z_{34} }, \quad 
I_{60}        = {1\over z_{21} z_{23} z_{45} } 
$$
\eqn\setInts{
I_{61}        = {1\over z_{21} z_{23} z_{34} }, \quad 
I_{62}        = {1\over z_{21} z_{23} z_{24} }, \quad 
I_{63}        = {1\over z_{21} z_{41} z_{23} } 
}

%***********************************
\subsec{Partial fraction identities}
%***********************************

It is possible to express $I_{35},{\ldots},I_{63}$ in
terms of $I_1,{\ldots},I_{34}$. To see this one uses the
partial fraction relations

\eqn\PFs{
{1\over z_{ji} z_{ki} } + {1\over z_{jk} z_{ji} } = {1\over z_{jk} z_{ki} },
}
and multiply each one of them by appropriate factors of ${1\over z_{ij}}$
to generate the following system of equations:
$$
      I_{37} - I_{14} + I_{8} = 0, \quad
    - I_{38} + I_{33} + I_{20} = 0, \quad
      I_{39} - I_{20} + I_{8} = 0, \quad
      I_{40} - I_{38} + I_{14} = 0, \quad
$$
$$
      I_{41} - I_{31} + I_{26} = 0, \quad
      I_{42} - I_{26} + I_{8} = 0, \quad
      I_{43} - I_{41} + I_{37} = 0, \quad
      I_{44} + I_{38} - I_{31} = 0, \quad
$$
$$
      I_{45} - I_{41} + I_{33} = 0, \quad
      I_{46} + I_{28} - I_{27} = 0, \quad
      I_{46} - I_{36} + I_{34} = 0, \quad
    - I_{46} + I_{35} + I_{29} = 0, \quad
$$
$$
      I_{47} - I_{43} + I_{40} = 0, \quad
      I_{47} - I_{45} + I_{39} = 0, \quad
      I_{47} - I_{44} + I_{42} = 0, \quad
    - I_{48} + I_{32} + I_{16} = 0, \quad
$$
$$
      I_{49} - I_{23} + I_{7} = 0, \quad
      I_{50} - I_{16} + I_{1} = 0, \quad
      I_{51} + I_{48} - I_{30} = 0, \quad
      I_{52} - I_{13} + I_{5} = 0, \quad
$$
$$
      I_{53} - I_{17} + I_{6} = 0, \quad
      I_{54} - I_{30} + I_{22} = 0, \quad
      I_{55} - I_{54} + I_{32} = 0, \quad
      I_{56} - I_{10} + I_{1} = 0, \quad
$$
$$
      I_{57} - I_{11} + I_{4} = 0, \quad
      I_{58} - I_{19} + I_{3} = 0, \quad
      I_{59} - I_{48} + I_{10} = 0, \quad
      I_{60} - I_{25} + I_{2} = 0, \quad
$$
$$
      I_{61} + I_{56} - I_{54} = 0, \quad
      I_{62} - I_{61} + I_{59} = 0, \quad
      I_{62} - I_{55} + I_{50} = 0, \quad
      I_{63} + I_{62} - I_{51} = 0, \quad
$$
\eqn\sysPFs{
      I_{63} - I_{22} + I_{1} = 0.
}
The solution is given by
$$
        I_{35} = I_{27} - I_{28} - I_{29},\quad
        I_{36} = I_{34} + I_{27} - I_{28},\quad
        I_{37} = I_{14} - I_{8},\quad
        I_{38} = I_{33} + I_{20}
$$
$$
        I_{39} = I_{20} - I_{8},\quad
        I_{40} = I_{33} + I_{20} - I_{14},\quad
        I_{41} = I_{31} - I_{26},\quad
        I_{42} =  I_{26} - I_{8}
$$
$$
        I_{43} = I_{31} - I_{26} - I_{14} + I_{8},\quad
        I_{44} = I_{31} - I_{33} - I_{20},\quad
        I_{45} = I_{31} - I_{26} - I_{33}
$$
$$
        I_{46} = I_{27} - I_{28},\quad
        I_{47} = I_{31} - I_{33} - I_{20} - I_{26} + I_{8},\quad
        I_{48} = I_{32} + I_{16},\quad
        I_{49} = I_{23} - I_{7}
$$
$$
        I_{50} = I_{16} - I_{1},\quad
        I_{51} = I_{30} - I_{32} - I_{16},\quad
        I_{52} = I_{13} - I_{5},\quad
        I_{53} = I_{17} - I_{6}
$$
$$
        I_{54} = I_{30} - I_{22},\quad
        I_{55} = I_{30} - I_{22} - I_{32},\quad
        I_{56} = I_{10} - I_{1},\quad
        I_{57} = I_{11} - I_{4},\quad
        I_{58} = I_{19} - I_{3}
$$
$$
        I_{59} = I_{16} - I_{10} + I_{32},\quad
        I_{60} = I_{25} - I_{2},\quad
        I_{61} = I_{30} - I_{22} - I_{10} + I_{1}
$$
\eqn\PFsolns{
        I_{62} = I_{1} - I_{16} - I_{22} + I_{30} - I_{32},\quad
        I_{63} = I_{22} - I_{1}
}

%**********************************
\subsec{Total derivative relations}
%**********************************

With the $SL(2,{\bf R})$ fixing choice of $(z_1,z_5,z_6)= (0,1,\infty)$, all the integrals $\cI_j$ share the
common factor of
$$
M(z_2,z_3,z_4)= z_2^{-s_{12}} z_3^{-s_{13}} z_4^{-s_{14}} (z_3-z_2)^{-s_{23}} (z_4-z_3)^{-s_{34}}
$$
\eqn\Mdef{
(z_4-z_2)^{-s_{24}}
     (1-z_2)^{-s_{25}} (1-z_3)^{-s_{35}} (1-z_4)^{-s_{45}}.
}
One can obtain additional relations among the integrals using the vanishing of
\eqn\totI{
\int dz_2 \int dz_3 \int dz_4 {\p\over \p z_I}\( 
{M \over z_{ij}z_{kl}}\) = 0, \quad i,j,k,l \neq I, \quad I=2,3,4
}
\eqn\totII{
\int dz_2 \int dz_3 \int dz_4 {\p\over \p z_I}\( 
{M \over z_{Ij} z_{kl}}\) = 0, \quad j,k,l \neq I, \quad I=2,3,4.
}
Equation \totI\ leads to 27 equations like
$$
 s_{12} \cI_{1} + s_{23}\cI_{22} + s_{24}\cI_{16} + s_{25}\cI_{5} = 0,
$$
$$
 s_{12}\cI_{2} + s_{23}\cI_{25} + s_{24}\cI_{17} + s_{25}\cI_{6} = 0,
$$
etc. Likewise, equation \totII\ gives rise to 18 relations,
$$
   (1+s_{34})\cI_{9} =
   (s_{13}+s_{23})\cI_{1}
   + s_{35}\cI_{4}
   - (s_{13} + s_{23})\cI_{10} 
   - s_{35}\cI_{11}
   - s_{23}\cI_{22}
   + s_{23}\cI_{30} 
$$
$$
  (1+s_{34})\cI_{12} =  
  s_{13}\cI_{5}
  + (s_{23} + s_{35})\cI_{8}
  - s_{13}\cI_{13}
  - (s_{23} + s_{35})\cI_{14}
  - s_{23}\cI_{26}
  + s_{23}\cI_{31}
$$
etc. It is straightforward to show that this system of equations
allow all the integrals to be rewritten in terms of a basis containing six elements
in agreement with the findings of~\StieOpr.

%*************************************************
\subsec{Cancelling the tachyon poles}
%*************************************************

Subsection 3.2 discusses the double pole integrals which by themselves introduce spurious tachyon poles. In particular, the four integrals in \doubleD\ are proportional to $\sim (1+t_2)^{-1}$ which is not at all obvious to cancel. This appendix explains the mechanisms leading to their cancellation.

Let us first of all plug in the explicit
superspace expressions for the kinematic the factors in \doubleD:
$$
 \langle L_{233434} V_1 V_5 V_6\rangle \cI_{27}
+ \langle L_{243434} V_1 V_5 V_6\rangle \cI_{28}
+ \langle L_{242434} V_1 V_5 V_6\rangle \cI_{29}
+ \langle L_{232334} V_1 V_5 V_6\rangle \cI_{36} =
$$
$$
 \langle \big[ (D_{34}(k^3\cdot A^2) - D_{24}(k^2\cdot A^3)) (1+s_{24}+s_{34})
+ D_{23}(k^2\cdot A^4) (1+s_{23}+s_{34})
\big]V_1 V_5 V_6\rangle \cI_{27}
$$
$$
+ \langle \big[(D_{24}(k^4\cdot A^3) - D_{34}(k^4\cdot A^2)) s_{23}
- D_{23}(k^3\cdot A^4)s_{24}\big] V_1 V_5 V_6\rangle \cI_{27}
$$
$$
+ \langle \big[(D_{34}(k^4\cdot A^2) - D_{23}(k^2\cdot A^4)) (1+s_{23}+s_{34})
+ D_{24}(k^2\cdot A^3) (1+s_{24}+s_{34}) \big] V_1 V_5 V_6\rangle \cI_{28}
$$
$$
+ \langle \big[ (D_{23}(k^3\cdot A^4) - D_{34}(k^3\cdot A^2)) s_{24}
- D_{24}(k^4\cdot A^3) s_{23} \big] V_1 V_5 V_6\rangle\cI_{28}
$$
\eqn\difficult{
+ \langle D_{24}\big[A^3\cdot (k^2+k^4)\big] V_1 V_5 V_6\rangle (1+s_{24}) {\cal I}_{29}
- \langle D_{23}\big[A^4\cdot (k^2+k^3)\big] V_1 V_5 V_6\rangle (1+s_{23}) {\cal I}_{36}
}
By virtue of the total derivative equations
$$
(1+s_{23})\cI_{36} = s_{24}(\cI_{28} - \cI_{27}) + {\cal R}_1, \quad
(1+s_{24})\cI_{29} = s_{23}(\cI_{28} - \cI_{27}) + {\cal R}_2
$$
where ${\cal R}_1$ and ${\cal R}_2$ denote integrals free of tachyonic poles
$$
{\cal R}_1 = - s_{12}\cI_{30} + s_{12}\cI_{22} + s_{12}\cI_{10} - s_{12}\cI_{1} - s_{25}\cI_{31} 
+ s_{25}\cI_{26} + s_{25}\cI_{14} - s_{25}\cI_{8}
$$
\eqn\regulars{
{\cal R}_2 = - s_{12}\cI_{32} - s_{12}\cI_{16} + s_{12}\cI_{10} - s_{25}\cI_{33} - s_{25}\cI_{20} 
+ s_{25}\cI_{14},
}
the RHS of \difficult\ becomes
$$
= \langle \big[ D_{23}(k^2\cdot A^4) - D_{24}(k^2\cdot A^3)\big] 
V_1 V_5 V_6\rangle (1+s_{23}+s_{24}+s_{34})(\cI_{27}-\cI_{28})
$$
$$
+ \langle D_{34}(k^3\cdot A^2) V_1 V_5 V_6\rangle\big[ (1+s_{24}+s_{34})\cI_{27} - s_{24}\cI_{28}\big]
+ \langle R_{234} V_1V_5V_6\rangle {\cal R}_1 
+ \langle R_{243} V_1V_5V_6\rangle {\cal R}_2
$$
\eqn\chuckberry{
+ \langle D_{34}(k^4\cdot A^2) V_1 V_5 V_6\rangle\big[ (1+s_{23}+s_{34})\cI_{28} - s_{23}\cI_{27}\big]
}
Furthermore, using
\eqn\twoeqs{
  (1+s_{24} + s_{34})\cI_{27} -s_{24}\cI_{28} = s_{14}\cI_{30} + s_{45}\cI_{31}, \quad
  (1+s_{23}+s_{34})\cI_{28} - s_{23}\cI_{27} = - s_{13}\cI_{32} - s_{35}\cI_{33}
}
and in particular their difference, which manifestly cancels the $(1+t_2)^{-1}$ tachyon pole
\eqn\regstq{
  (1+s_{23} + s_{24} + s_{34})(\cI_{27}-\cI_{28}) = s_{14}\cI_{30} + s_{45}\cI_{31}
 + s_{13}\cI_{32}  + s_{35}\cI_{33}
}
allows \chuckberry\ to be rewritten in terms of unproblematic single-pole integrals:
$$
\langle R_{234}V_1V_5V_6\rangle {\cal R}_1 + \langle R_{243} V_1V_5V_6\rangle {\cal R}_2
$$
$$
 + \langle \big[ D_{34}(k^3\cdot A^2) + D_{23}(k^2\cdot A^4) 
 - D_{24}(k^2\cdot A^3)\big]V_1V_5V_6\rangle (s_{14}\cI_{30}+s_{45}\cI_{31})
$$
$$
- \langle \big[ D_{34}(k^4\cdot A^2) - D_{23}(k^2\cdot A^4) + D_{24}(k^2\cdot A^3)\big]V_1V_5V_6\rangle
(s_{13}\cI_{32}+s_{35}\cI_{33}).
$$

%*************************************************
\subsec{Momentum expansion of the $F_i$ integrals}
%*************************************************

Here we list the momentum expansion of the $F_i$ integrals in our end result \simpleSpt\ 
for the six point amplitude. The first 
three integrals $F_{1,2,3}$ (and their parity images) multiply superfield kinematics 
of type $\langle T_{ij} T_{kl} T_{mn} \rangle$. Their field theory contribution is 
of order ${\cal O}(s_{ij}^{-3}) = {\cal O}(k^{-6})$, then the first superstring 
corrections ${\cal O}(s_{ij}^{-1}) = {\cal O}(k^{-2})$ and 
${\cal O}(s_{ij}^{0}) = {\cal O}(k^{0})$ are multiplied by the transcendental 
numbers $\zeta(2)$ and $\zeta(3)$ respectively\foot{The field theory- and $\zeta(2)$ parts of the ${\cal I}_j$ are 
odd in the $s_i,t_i$ whereas the $\zeta(3)$ correction is even. That is why one has to be careful
about the sign convention of the Mandelstam variables. The $s_i,t_i$ in the present paper as well
as the references \refs{\mafrabcj,\FTAmps}  have to be replaced by $-s_i, -t_i$ for comparison 
with \refs{\StieOpr,\stieMultiGluon,\stieS}.}. More precisely, only $F_1$ has a 
nonzero field theory limit, and the superstring corrections $\sim \zeta(2),\zeta(3)$ have 
no more than two poles at the same time reflecting the fact that they represent contact 
interactions. $F_3$ is even limited to single poles.

The remaining integrals $F_4$ to $F_{10}$ associated 
with $\langle T_{ijk} E_{lmn} \rangle$ kinematics have an additional power of Mandelstam 
invariants in their field theory-, $\zeta(2)$- and $\zeta(3)$
terms, in order to compensate the $s_{ij}^{-1}$ within the definition of $E_{lmn}$. 
Once again, we find a hierarchy in their pole structure: The first three integrals 
$F_4,F_5,F_6$ have two kinematic poles in their field theory limit and mostly single 
poles in their $\zeta(2), \zeta(3)$ corrections. The few exceptional double poles in 
these higher order contributions -- say ${s_4 s_5 \zeta(2) \over s_1 t_1}$ in $F_4$ -- 
are decorated by numerators which cancel both poles in the associated $E_{ijk}$, 
e.g. $s_4 s_5 E_{456} = s_5 T_{45} V_6 + s_4 V_4 T_{56}$. 
The integrals $F_{7},F_{8}$ have single 
poles only from the beginning, and $F_9, F_{10}$ are even completely regular 
and start at $\zeta(3)$.

There is an infinite tower of higher order corrections in the momenta, i.e. higher 
orders in $\alpha'$,  along with multi--zeta values (MZVs) which we do not display here.
$$\eqalign{ 
&F_1 = - \int dz_2 \int dz_3 \int dz_4 \prod_{i<j}^6 { |z_{ij}|^{-s_{ij}} \over s_6} \Big( 
       {s_{12} \over z_{21} z_{23} z_{45}} + {s_{13} \over z_{23} z_{31} z_{45}} \Big)\cr
&= {1 \over s_2 s_4 s_6} 
 + \zeta(2)\ \Big( 
     { t_1 \over s_2 s_6} - {t_1 \over s_2 s_4} - {s_5 \over s_2 s_6} - {t_3 \over s_4 s_6} 
  +  {t_1 \over s_4 s_6} + {1 \over s_4} - {s_2 \over s_4 s_6} - {s_1 \over s_4 s_6} 
\Big)\cr
&+ \zeta(3)\ \Big( - {t_1^2 \over s_2 s_4} - {s_6 t_1 \over s_2 s_4} 
+ {2 t_1 t_2 \over s_2 s_6} + {t_1^2 \over s_2 s_6}
 - {s_5^2 \over s_2 s_6} + {s_4 t_1 \over s_2 s_6}
- {s_4 s_5 \over s_2 s_6} - {t_3^2 \over s_4 s_6}
+ {t_1 t_3 \over s_4 s_6}
+ {t_1^2 \over s_4 s_6} \cr
&+ {t_1 \over s_4} + {s_6 \over s_4}
- {2 t_2 \over s_6} - {2 t_1 \over s_6}
 + {2 s_3 \over s_6} - {2 s_2 t_3 \over s_4 s_6}
 + {2 s_2 \over s_6} - {s_2^2 \over s_4 s_6}
- {s_1 t_3 \over s_4 s_6}
+ {s_1 \over s_4}
- {2 s_1 s_2 \over s_4 s_6} - {s_1^2 \over s_4 s_6} \Big)+\ldots\ ,
}
$$
$$
\eqalign{
F_2 &= - \int dz_2 \int dz_3 \int dz_4 \prod_{i<j}^6 { |z_{ij}|^{-s_{ij}} \over s_6} \Big( 
 {s_{14} \over z_{25} z_{34} z_{41}} + {s_{13} \over z_{25} z_{34} z_{31}} \Big)\cr
 &= \zeta(2)\ \Big( - {t_2 \over s_3 s_6}  + {s_5 \over s_3 s_6} - {s_1 \over s_3 s_6} \Big) 
+  \zeta(3)\ \Big( - {t_2 t_3 \over s_3 s_6} - {t_2^2 \over s_3 s_6} 
- {t_2 \over s_3} + {s_5 t_3 \over s_3 s_6} + {s_5 \over s_3} \cr
&+ {s_5^2 \over s_3 s_6} + {t_2 \over s_6} - {t_1 \over s_6} - {s_5 \over s_6} + {s_2 \over s_6}
- {s_1 t_3 \over s_3 s_6} - {2 s_1 t_2 \over s_3 s_6} - {s_1 \over s_3} 
+ {2 s_1 \over s_6} - {s_1^2 \over s_3 s_6} \Big)+\ldots\ ,\cr\crr
%%%%%
F_3 &= - \int dz_2 \int dz_3 \int dz_4 \prod_{i<j}^6 { |z_{ij}|^{-s_{ij}} \over s_6} \Big( 
 {s_{14} \over z_{24} z_{41} z_{35}} + {s_{12} \over z_{24} z_{21} z_{35}} 
 \Big)\cr
&= {\zeta(2) \over s_6} + \zeta(3)\  \Big( { t_3 \over s_6} + {2 t_2 \over s_6} 
+ {3 t_1 \over s_6} - {s_5 \over s_6} + {s_4 \over s_6} + {s_3 \over s_6} 
- {2s_2 \over s_6} - {2 s_1 \over s_6} \Big)+\ldots\ ,\cr\crr
%%%%%
F_4 &= \int dz_2 \int dz_3 \int dz_4 \prod_{i<j}^6  |z_{ij}|^{-s_{ij}} { s_{45}  \over z_{21} z_{31} z_{45} }\cr
& = {1 \over s_1 t_1} + \zeta(2)\ \Big( {s_4 \over s_1} - {t_3 \over s_1} - {s_4 s_5 \over s_1 t_1} 
  - {s_2 \over t_1} \Big)  + \zeta(3)\ \Big( {s_4 t_1 \over s_1} - {t_3^2 \over s_1} - {t_1 t_3 \over s_1} 
  - {s_4 s_5^2 \over s_1 t_1} \cr
& + {s_4^2 \over s_1} - {s_4^2 s_5 \over s_1 t_1} + {2 s_3 s_4 \over s_1} - {s_2^2 \over t_1} 
  - {s_1 s_2 \over t_1} - s_2 + s_6 - t_3 \Big)+\ldots\ ,\cr\crr
%%%%%
F_5 & =\int dz_2 \int dz_3 \int dz_4 \prod_{i<j}^6  |z_{ij}|^{-s_{ij}} {1 \over z_{23} z_{24}} 
       \Big( {s_{12} \over z_{21}} + {s_{13} \over z_{31}}  + {s_{14} \over z_{41}} \Big)\cr
& = - {1 \over s_2 t_2} + \zeta(2)\ \Big( {t_2 \over s_2} + {t_1 \over s_2} - 1 - {s_6 \over s_2} 
    - {s_5 \over s_2} + {s_5 s_6 \over s_2 t_2} + {s_3 \over t_2} + {s_4 \over s_2} \Big) \cr
& + \zeta(3)\ \Big( {t_1^2 \over s_2} + {2 t_1 t_2 \over s_2} + {t_2^2 \over s_2} 
  - {s_5^2 \over s_2} - {s_6 t_1 \over s_2}- {s_5 s_6 \over s_2} + {s_5^2 s_6 \over s_2 t_2} 
  - {s_6^2 \over s_2} + {s_5 s_6^2 \over s_2 t_2} - 2 t_2  \cr
& - t_3 + s_6 - 2 s_1 - s_2 + s_3 + {s_2 s_3 \over t_2} + {s_3^2 \over t_2} + {s_4 t_1 \over s_2} 
  + {2 s_4 t_2 \over s_2} - {s_4 s_5 \over s_2} + {s_4^2\over s_2} \Big) +\ldots\ ,\cr\crr
%%%%
F_6 &= \int dz_2 \int dz_3 \int dz_4 \prod_{i<j}^6  |z_{ij}|^{-s_{ij}} 
{ s_{12}  \over z_{21} z_{34} z_{35} }\cr
&= -{1 \over s_3 t_3} + \zeta(2)\ \Big( {s_5 \over s_3} - {s_1 \over s_3} + {s_1 s_6 \over s_3 t_3} 
+ {s_4 \over t_3} - 1 \Big) 
+ \zeta(3)\ \Big( {s_5 t_3 \over s_3} + {s_5^2 \over s_3} - {2 s_1 t_2 \over s_3}  \cr
& - {s_1 t_3 \over s_3} + {s_1 s_6^2 \over s_3 t_3} - {s_1^2 \over s_3} +  {s_1^2 s_6 \over s_3 t_3} - 2 t_1 
- t_3 + 2 s_1 -s_3 + {s_3 s_4 \over t_3} + {s_4^2 \over t_3} \Big)+\ldots\ ,\cr\crr
F_7 &=\int dz_2 \int dz_3 \int dz_4 \prod_{i<j}^6  |z_{ij}|^{-s_{ij}} 
{ s_{14}  \over z_{23} z_{25} z_{41} }\cr
&= -\zeta(2)\  {s_{14} \over s_2} -\zeta(3)\ {s_{14}  \over s_2} \Big( t_1 + t_2 + s_4 + s_5 + s_6 \Big) 
+ \zeta(3)\ s_{14}+\ldots\ ,\cr\crr
}
$$
$$
\eqalign{
F_8 &= \int dz_2 \int dz_3 \int dz_4 \prod_{i<j}^6  |z_{ij}|^{-s_{ij}} {  s_{35}  \over z_{21}  z_{35} z_{41}}\cr
&=\zeta(2)\ { s_{35} \over s_1}+\zeta(3)\ { s_{35} \over s_1}\ (t_1 + t_3 + s_3 + s_4 + s_5)+\ldots\ ,\cr\crr
F_9 &= \int dz_2 \int dz_3 \int dz_4 \prod_{i<j}^6  |z_{ij}|^{-s_{ij}} {  s_{14}  \over z_{25}  z_{35} z_{41} } =
\zeta(3)\ s_{14}+\ldots\ ,\cr\crr
F_{10} &= \int dz_2 \int dz_3 \int dz_4 \prod_{i<j}^6  |z_{ij}|^{-s_{ij}} 
 { s_{35}  \over z_{24} z_{35} z_{41} } = 2\ \zeta(3)\ s_{35}+\ldots\ .
 }
 $$

%****************************************************************
\appendix{C}{Six--point subamplitudes and extended BCJ relations}
%****************************************************************

In this Appendix we display the parametrization of the remaining $21$ partial subamplitudes:
$$\eqalign{
A(1,2,4,3,5,6)&=
- {n_{1}(2345)\over s_{12} s_{34} s_{56}}
+ {n_{2}(2435)\over s_{24} s_{35} s_{124}}
+ {n_{3}(2435)\over s_{12} s_{35} s_{124}}
+ {n_{4}(2435)\over s_{12} s_{56} s_{124}}
+ {n_{5}(2435)\over s_{24} s_{56} s_{124}}\cr\crr
&+ {n_{6}(2435)\over s_{24} s_{56} s_{156}}
- {n_{7}(2345)\over s_{34} s_{56} s_{156}}
+ {n'_1(2435)\over s_{16} s_{24} s_{35}}
- {n'_2(2345)\over s_{12} s_{34} s_{126}}
- {n'_2(2354)\over s_{12} s_{35} s_{126}}\cr\crr
&- {n'_5(2345)\over s_{16} s_{34} s_{126}}
- {n'_5(2354)\over s_{16} s_{35} s_{126}}
- {n'_6(2345)\over s_{16} s_{34} s_{156}}
+ {n'_7(2435)\over s_{16} s_{24} s_{156}}\ ,\cr\cr
A(1,2,4,5,3,6)&=
  {n_{1}(2453)\over s_{12} s_{36} s_{45}}
- {n_{2}(2435)\over s_{24} s_{35} s_{124}}
- {n_{3}(2435)\over s_{12} s_{35} s_{124}}
+ {n_{4}(2453)\over s_{12} s_{36} s_{124}}
+ {n_{5}(2453)\over s_{24} s_{36} s_{124}}\cr\crr
&+ {n_{6}(2453)\over s_{24} s_{36} s_{136}}
+ {n_{7}(2453)\over s_{36} s_{45} s_{136}}
- {n'_1(2435)\over s_{16} s_{24} s_{35}}
+ {n'_2(2354)\over s_{12} s_{35} s_{126}}
- {n'_3(2345)\over s_{12} s_{45} s_{126}}\cr
&- {n'_4(2345)\over s_{16} s_{45} s_{126}}
+ {n'_5(2354)\over s_{16} s_{35} s_{126}}
+ {n'_6(2453)\over s_{16} s_{45} s_{136}}
+ {n'_7(2453)\over s_{16} s_{24} s_{136}}\ ,\cr\cr
A(1,2,5,3,4,6)&=
- {n_{1}(2354)\over s_{12} s_{35} s_{46}}
+ {n_{2}(2534)\over s_{25} s_{34} s_{125}}
+ {n_{3}(2534)\over s_{12} s_{34} s_{125}}
+ {n_{4}(2534)\over s_{12} s_{46} s_{125}}
+ {n_{5}(2534)\over s_{25} s_{46} s_{125}}\cr
&+ {n_{6}(2534)\over s_{25} s_{46} s_{146}}
- {n_{7}(2354)\over s_{35} s_{46} s_{146}}
+ {n'_1(2534)\over s_{16} s_{25} s_{34}}
- {n'_2(2345)\over s_{12} s_{34} s_{126}}
- {n'_2(2354)\over s_{12} s_{35} s_{126}}\cr
&- {n'_5(2345)\over s_{16} s_{34} s_{126}}
- {n'_5(2354)\over s_{16} s_{35} s_{126}}
- {n'_6(2354)\over s_{16} s_{35} s_{146}}
+ {n'_7(2534)\over s_{16} s_{25} s_{146}}\ ,}
$$
%%%%%%%%%%%%%%%%%%%%%%%%%%%%%%%%%%%%%%%%%%%%%%%%%%%%
$$\eqalign{
A(1,2,5,4,3,6)&=
- {n_{1}(2453)\over s_{12} s_{36} s_{45}}
- {n_{2}(2534)\over s_{25} s_{34} s_{125}}
- {n_{3}(2534)\over s_{12} s_{34} s_{125}}
+ {n_{4}(2543)\over s_{12} s_{36} s_{125}}
+ {n_{5}(2543)\over s_{25} s_{36} s_{125}}\cr
&+ {n_{6}(2543)\over s_{25} s_{36} s_{136}}
- {n_{7}(2453)\over s_{36} s_{45} s_{136}}
- {n'_1(2534)\over s_{16} s_{25} s_{34}}
+ {n'_2(2345)\over s_{12} s_{34} s_{126}}
+ {n'_3(2345)\over s_{12} s_{45} s_{126}}\cr\crr
&+ {n'_4(2345)\over s_{16} s_{45} s_{126}}
+ {n'_5(2345)\over s_{16} s_{34} s_{126}}
- {n'_6(2453)\over s_{16} s_{45} s_{136}}
+ {n'_7(2543)\over s_{16} s_{25} s_{136}}\ ,\cr\cr
A(1,3,2,4,5,6)&=  {n_{1}(3245)\over s_{13} s_{24} s_{56}}
- {n_{2}(2345)\over s_{23} s_{45} s_{123}}
+ {n_{3}(3245)\over s_{13} s_{45} s_{123}}
+ {n_{4}(3245)\over s_{13} s_{56} s_{123}}
- {n_{5}(2345)\over s_{23} s_{56} s_{123}}\cr
&- {n_{6}(2345)\over s_{23} s_{56} s_{156}}
- {n_{6}(2435)\over s_{24} s_{56} s_{156}}
- {n'_1(2345)\over s_{16} s_{23} s_{45}}
+ {n'_2(3245)\over s_{13} s_{24} s_{136}}
+ {n'_3(3245)\over s_{13} s_{45} s_{136}}\cr\crr
&- {n'_6(2453)\over s_{16} s_{45} s_{136}}
- {n'_7(2345)\over s_{16} s_{23} s_{156}}
- {n'_7(2435)\over s_{16} s_{24} s_{156}}
- {n'_7(2453)\over s_{16} s_{24} s_{136}}\ ,\cr\cr
A(1,3,2,5,4,6)&=
  {n_{1}(3254)\over s_{13} s_{25} s_{46}}
+ {n_{2}(2345)\over s_{23} s_{45} s_{123}}
- {n_{3}(3245)\over s_{13} s_{45} s_{123}}
+ {n_{4}(3254)\over s_{13} s_{46} s_{123}}
- {n_{5}(2354)\over s_{23} s_{46} s_{123}}\cr\crr
&- {n_{6}(2354)\over s_{23} s_{46} s_{146}}
- {n_{6}(2534)\over s_{25} s_{46} s_{146}}
+ {n'_1(2345)\over s_{16} s_{23} s_{45}}
+ {n'_2(3254)\over s_{13} s_{25} s_{136}}
- {n'_3(3245)\over s_{13} s_{45} s_{136}}\cr\crr
&+ {n'_6(2453)\over s_{16} s_{45} s_{136}}
- {n'_7(2354)\over s_{16} s_{23} s_{146}}
- {n'_7(2534)\over s_{16} s_{25} s_{146}}
- {n'_7(2543)\over s_{16} s_{25} s_{136}}\ ,\cr\cr
A(1,3,4,2,5,6)&=
- {n_{1}(3245)\over s_{13} s_{24} s_{56}}
+ {n_{2}(3425)\over s_{25} s_{34} s_{134}}
+ {n_{3}(3425)\over s_{13} s_{25} s_{134}}
+ {n_{4}(3425)\over s_{13} s_{56} s_{134}}
+ {n_{5}(3425)\over s_{34} s_{56} s_{134}}\cr\crr
&+ {n_{6}(2435)\over s_{24} s_{56} s_{156}}
- {n_{7}(2345)\over s_{34} s_{56} s_{156}}
- {n'_1(2534)\over s_{16} s_{25} s_{34}}
- {n'_2(3245)\over s_{13} s_{24} s_{136}}
- {n'_2(3254)\over s_{13} s_{25} s_{136}}\cr\crr
&- {n'_6(2345)\over s_{16} s_{34} s_{156}}
+ {n'_7(2435)\over s_{16} s_{24} s_{156}}
+ {n'_7(2453)\over s_{16} s_{24} s_{136}}
+ {n'_7(2543)\over s_{16} s_{25} s_{136}}\ ,\cr\cr
A(1,3,4,5,2,6)&=
  {n_{1}(3452)\over s_{13} s_{26} s_{45}}
- {n_{2}(3425)\over s_{25} s_{34} s_{134}}
- {n_{3}(3425)\over s_{13} s_{25} s_{134}}
+ {n_{4}(3452)\over s_{13} s_{26} s_{134}}
+ {n_{5}(3452)\over s_{26} s_{34} s_{134}}\cr\crr
&+ {n_{6}(3452)\over s_{26} s_{34} s_{126}}
+ {n_{7}(3452)\over s_{26} s_{45} s_{126}}
+ {n'_1(2534)\over s_{16} s_{25} s_{34}}
+ {n'_2(3254)\over s_{13} s_{25} s_{136}}
- {n'_3(3245)\over s_{13} s_{45} s_{136}}\cr\crr
&- {n'_4(2345)\over s_{16} s_{45} s_{126}}
- {n'_5(2345)\over s_{16} s_{34} s_{126}}
+ {n'_6(2453)\over s_{16} s_{45} s_{136}}
- {n'_7(2543)\over s_{16} s_{25} s_{136}}\ ,}$$
%%%%%%%%%%%%%%%%%%%%%%%%%%%%%%%%%%%%%%%%%%%%%%%%%%%%%%
$$\eqalign{
A(1,3,5,2,4,6)&=
- {n_{1}(3254)\over s_{13} s_{25} s_{46}}
+ {n_{2}(3524)\over s_{24} s_{35} s_{135}}
+ {n_{3}(3524)\over s_{13} s_{24} s_{135}}
+ {n_{4}(3524)\over s_{13} s_{46} s_{135}}
+ {n_{5}(3524)\over s_{35} s_{46} s_{135}}\cr\crr
&+ {n_{6}(2534)\over s_{25} s_{46} s_{146}}
- {n_{7}(2354)\over s_{35} s_{46} s_{146}}
- {n'_1(2435)\over s_{16} s_{24} s_{35}}
- {n'_2(3245)\over s_{13} s_{24} s_{136}}
- {n'_2(3254)\over s_{13} s_{25} s_{136}}\cr\crr
&- {n'_6(2354)\over s_{16} s_{35} s_{146}}
+ {n'_7(2453)\over s_{16} s_{24} s_{136}}
+ {n'_7(2534)\over s_{16} s_{25} s_{146}}
+ {n'_7(2543)\over s_{16} s_{25} s_{136}}\ ,\cr\cr
%%%%%%%%%%%%%%%%%%%%%%%%%%%%%%%%%%%%%%%%%%%%%%%%%%%%%%%%%%%%%
A(1,3,5,4,2,6)&=
- {n_{1}(3452)\over s_{13} s_{26} s_{45}}
- {n_{2}(3524)\over s_{24} s_{35} s_{135}}
- {n_{3}(3524)\over s_{13} s_{24} s_{135}}
+ {n_{4}(3542)\over s_{13} s_{26} s_{135}}
+ {n_{5}(3542)\over s_{26} s_{35} s_{135}}\cr\crr
&+ {n_{6}(3542)\over s_{26} s_{35} s_{126}}
- {n_{7}(3452)\over s_{26} s_{45} s_{126}}
+ {n'_1(2435)\over s_{16} s_{24} s_{35}}
+ {n'_2(3245)\over s_{13} s_{24} s_{136}}
+ {n'_3(3245)\over s_{13} s_{45} s_{136}}\cr\crr
&+ {n'_4(2345)\over s_{16} s_{45} s_{126}}
- {n'_5(2354)\over s_{16} s_{35} s_{126}}
- {n'_6(2453)\over s_{16} s_{45} s_{136}}
- {n'_7(2453)\over s_{16} s_{24} s_{136}}\ ,\cr\cr
A(1,4,2,3,5,6)&=
  {n_{1}(4235)\over s_{14} s_{23} s_{56}}
- {n_{2}(2435)\over s_{24} s_{35} s_{124}}
+ {n_{3}(4235)\over s_{14} s_{35} s_{124}}
+ {n_{4}(4235)\over s_{14} s_{56} s_{124}}
- {n_{5}(2435)\over s_{24} s_{56} s_{124}}\cr\crr
&- {n_{6}(2345)\over s_{23} s_{56} s_{156}}
- {n_{6}(2435)\over s_{24} s_{56} s_{156}}
- {n'_1(2435)\over s_{16} s_{24} s_{35}}
+ {n'_2(4235)\over s_{14} s_{23} s_{146}}
+ {n'_3(4235)\over s_{14} s_{35} s_{146}}\cr\crr
&- {n'_6(2354)\over s_{16} s_{35} s_{146}}
- {n'_7(2345)\over s_{16} s_{23} s_{156}}
- {n'_7(2354)\over s_{16} s_{23} s_{146}}
- {n'_7(2435)\over s_{16} s_{24} s_{156}}\ ,\cr\cr
A(1,4,2,5,3,6)&=
  {n_{1}(4253)\over s_{14} s_{25} s_{36}}
+ {n_{2}(2435)\over s_{24} s_{35} s_{124}}
- {n_{3}(4235)\over s_{14} s_{35} s_{124}}
+ {n_{4}(4253)\over s_{14} s_{36} s_{124}}
- {n_{5}(2453)\over s_{24} s_{36} s_{124}}\cr\crr
&- {n_{6}(2453)\over s_{24} s_{36} s_{136}}
- {n_{6}(2543)\over s_{25} s_{36} s_{136}}
+ {n'_1(2435)\over s_{16} s_{24} s_{35}}
+ {n'_2(4253)\over s_{14} s_{25} s_{146}}
- {n'_3(4235)\over s_{14} s_{35} s_{146}}\cr\crr
&+ {n'_6(2354)\over s_{16} s_{35} s_{146}}
- {n'_7(2453)\over s_{16} s_{24} s_{136}}
- {n'_7(2534)\over s_{16} s_{25} s_{146}}
- {n'_7(2543)\over s_{16} s_{25} s_{136}}\ ,\cr\cr
A(1,4,3,2,5,6)&=
- {n_{1}(4235)\over s_{14} s_{23} s_{56}}
- {n_{2}(3425)\over s_{25} s_{34} s_{134}}
+ {n_{3}(4325)\over s_{14} s_{25} s_{134}}
+ {n_{4}(4325)\over s_{14} s_{56} s_{134}}
- {n_{5}(3425)\over s_{34} s_{56} s_{134}}\cr\crr
&+ {n_{6}(2345)\over s_{23} s_{56} s_{156}}
+ {n_{7}(2345)\over s_{34} s_{56} s_{156}}
+ {n'_1(2534)\over s_{16} s_{25} s_{34}}
- {n'_2(4235)\over s_{14} s_{23} s_{146}}
- {n'_2(4253)\over s_{14} s_{25} s_{146}}\cr\crr
&+ {n'_6(2345)\over s_{16} s_{34} s_{156}}
+ {n'_7(2345)\over s_{16} s_{23} s_{156}}
+ {n'_7(2354)\over s_{16} s_{23} s_{146}}
+ {n'_7(2534)\over s_{16} s_{25} s_{146}}\ ,}$$
%%%%%%%%%%%%%%%%%%%%%%%%%%%%%%%%%%%%%%%%%%%%%%%%%%%%
$$\eqalign{
A(1,4,3,5,2,6)&=
  {n_{1}(4352)\over s_{14} s_{26} s_{35}}
+ {n_{2}(3425)\over s_{25} s_{34} s_{134}}
- {n_{3}(4325)\over s_{14} s_{25} s_{134}}
+ {n_{4}(4352)\over s_{14} s_{26} s_{134}}
- {n_{5}(3452)\over s_{26} s_{34} s_{134}}\cr\crr
&- {n_{6}(3452)\over s_{26} s_{34} s_{126}}
- {n_{6}(3542)\over s_{26} s_{35} s_{126}}
- {n'_1(2534)\over s_{16} s_{25} s_{34}}
+ {n'_2(4253)\over s_{14} s_{25} s_{146}}
- {n'_3(4235)\over s_{14} s_{35} s_{146}}\cr\crr
&+ {n'_5(2345)\over s_{16} s_{34} s_{126}}
+ {n'_5(2354)\over s_{16} s_{35} s_{126}}
+ {n'_6(2354)\over s_{16} s_{35} s_{146}}
- {n'_7(2534)\over s_{16} s_{25} s_{146}}\ ,\cr\cr
A(1,4,5,2,3,6)&=
- {n_{1}(4253)\over s_{14} s_{25} s_{36}}
+ {n_{2}(4523)\over s_{23} s_{45} s_{145}}
+ {n_{3}(4523)\over s_{14} s_{23} s_{145}}
+ {n_{4}(4523)\over s_{14} s_{36} s_{145}}
+ {n_{5}(4523)\over s_{36} s_{45} s_{145}}\cr\crr
&+ {n_{6}(2543)\over s_{25} s_{36} s_{136}}
- {n_{7}(2453)\over s_{36} s_{45} s_{136}}
- {n'_1(2345)\over s_{16} s_{23} s_{45}}
- {n'_2(4235)\over s_{14} s_{23} s_{146}}
- {n'_2(4253)\over s_{14} s_{25} s_{146}}\cr\crr
&- {n'_6(2453)\over s_{16} s_{45} s_{136}}
+ {n'_7(2354)\over s_{16} s_{23} s_{146}}
+ {n'_7(2534)\over s_{16} s_{25} s_{146}}
+ {n'_7(2543)\over s_{16} s_{25} s_{136}}\ ,\cr\cr
A(1,4,5,3,2,6)&=
- {n_{1}(4352)\over s_{14} s_{26} s_{35}}
- {n_{2}(4523)\over s_{23} s_{45} s_{145}}
- {n_{3}(4523)\over s_{14} s_{23} s_{145}}
+ {n_{4}(4532)\over s_{14} s_{26} s_{145}}
+ {n_{5}(4532)\over s_{26} s_{45} s_{145}}\cr\crr
&+ {n_{6}(3542)\over s_{26} s_{35} s_{126}}
- {n_{7}(3452)\over s_{26} s_{45} s_{126}}
+ {n'_1(2345)\over s_{16} s_{23} s_{45}}
+ {n'_2(4235)\over s_{14} s_{23} s_{146}}
+ {n'_3(4235)\over s_{14} s_{35} s_{146}}\cr\crr
&+ {n'_4(2345)\over s_{16} s_{45} s_{126}}
- {n'_5(2354)\over s_{16} s_{35} s_{126}}
- {n'_6(2354)\over s_{16} s_{35} s_{146}}
- {n'_7(2354)\over s_{16} s_{23} s_{146}}\ ,\cr\cr
A(1,5,2,3,4,6)&=
  {n_{1}(5234)\over s_{15} s_{23} s_{46}}
- {n_{2}(2534)\over s_{25} s_{34} s_{125}}
+ {n_{3}(5234)\over s_{15} s_{34} s_{125}}
+ {n_{4}(5234)\over s_{15} s_{46} s_{125}}
- {n_{5}(2534)\over s_{25} s_{46} s_{125}}\cr\crr
&- {n_{6}(2354)\over s_{23} s_{46} s_{146}}
- {n_{6}(2534)\over s_{25} s_{46} s_{146}}
- {n'_1(2534)\over s_{16} s_{25} s_{34}}
+ {n'_2(5234)\over s_{15} s_{23} s_{156}}
+ {n'_3(5234)\over s_{15} s_{34} s_{156}}\cr\crr
&- {n'_6(2345)\over s_{16} s_{34} s_{156}}
- {n'_7(2345)\over s_{16} s_{23} s_{156}}
- {n'_7(2354)\over s_{16} s_{23} s_{146}}
- {n'_7(2534)\over s_{16} s_{25} s_{146}}\ ,\cr\cr
A(1,5,2,4,3,6)&=
  {n_{1}(5243)\over s_{15} s_{24} s_{36}}
+ {n_{2}(2534)\over s_{25} s_{34} s_{125}}
- {n_{3}(5234)\over s_{15} s_{34} s_{125}}
+ {n_{4}(5243)\over s_{15} s_{36} s_{125}}
- {n_{5}(2543)\over s_{25} s_{36} s_{125}}\cr\crr
&- {n_{6}(2453)\over s_{24} s_{36} s_{136}}
- {n_{6}(2543)\over s_{25} s_{36} s_{136}}
+ {n'_1(2534)\over s_{16} s_{25} s_{34}}
+ {n'_2(5243)\over s_{15} s_{24} s_{156}}
- {n'_3(5234)\over s_{15} s_{34} s_{156}}\cr\crr
&+ {n'_6(2345)\over s_{16} s_{34} s_{156}}
- {n'_7(2435)\over s_{16} s_{24} s_{156}}
- {n'_7(2453)\over s_{16} s_{24} s_{136}}
- {n'_7(2543)\over s_{16} s_{25} s_{136}}\ ,}$$
%%%%%%%%%%%%%%%%%%%%%%%%%%%%%%%%%%%%%%%%%%%%%%%%%%%%%
\eqn\AAnsatz{\eqalign{
A(1,5,3,2,4,6)&=
- {n_{1}(5234)\over s_{15} s_{23} s_{46}}
- {n_{2}(3524)\over s_{24} s_{35} s_{135}}
+ {n_{3}(5324)\over s_{15} s_{24} s_{135}}
+ {n_{4}(5324)\over s_{15} s_{46} s_{135}}
- {n_{5}(3524)\over s_{35} s_{46} s_{135}}\cr\crr
&+ {n_{6}(2354)\over s_{23} s_{46} s_{146}}
+ {n_{7}(2354)\over s_{35} s_{46} s_{146}}
+ {n'_1(2435)\over s_{16} s_{24} s_{35}}
- {n'_2(5234)\over s_{15} s_{23} s_{156}}
- {n'_2(5243)\over s_{15} s_{24} s_{156}}\cr\crr
&+ {n'_6(2354)\over s_{16} s_{35} s_{146}}
+ {n'_7(2345)\over s_{16} s_{23} s_{156}}
+ {n'_7(2354)\over s_{16} s_{23} s_{146}}
+ {n'_7(2435)\over s_{16} s_{24} s_{156}}\ ,\cr\cr
A(1,5,3,4,2,6)&=
  {n_{1}(5342)\over s_{15} s_{26} s_{34}}
+ {n_{2}(3524)\over s_{24} s_{35} s_{135}}
- {n_{3}(5324)\over s_{15} s_{24} s_{135}}
+ {n_{4}(5342)\over s_{15} s_{26} s_{135}}
- {n_{5}(3542)\over s_{26} s_{35} s_{135}}\cr\crr
&- {n_{6}(3452)\over s_{26} s_{34} s_{126}}
- {n_{6}(3542)\over s_{26} s_{35} s_{126}}
- {n'_1(2435)\over s_{16} s_{24} s_{35}}
+ {n'_2(5243)\over s_{15} s_{24} s_{156}}
- {n'_3(5234)\over s_{15} s_{34} s_{156}}\cr\crr
&+ {n'_5(2345)\over s_{16} s_{34} s_{126}}
+ {n'_5(2354)\over s_{16} s_{35} s_{126}}
+ {n'_6(2345)\over s_{16} s_{34} s_{156}}
- {n'_7(2435)\over s_{16} s_{24} s_{156}}\ ,\cr\cr
A(1,5,4,2,3,6)&=
- {n_{1}(5243)\over s_{15} s_{24} s_{36}}
- {n_{2}(4523)\over s_{23} s_{45} s_{145}}
+ {n_{3}(5423)\over s_{15} s_{23} s_{145}}
+ {n_{4}(5423)\over s_{15} s_{36} s_{145}}
- {n_{5}(4523)\over s_{36} s_{45} s_{145}}\cr\crr
&+ {n_{6}(2453)\over s_{24} s_{36} s_{136}}
+ {n_{7}(2453)\over s_{36} s_{45} s_{136}}
+ {n'_1(2345)\over s_{16} s_{23} s_{45}}
- {n'_2(5234)\over s_{15} s_{23} s_{156}}
- {n'_2(5243)\over s_{15} s_{24} s_{156}}\cr\crr
&+ {n'_6(2453)\over s_{16} s_{45} s_{136}}
+ {n'_7(2345)\over s_{16} s_{23} s_{156}}
+ {n'_7(2435)\over s_{16} s_{24} s_{156}}
+ {n'_7(2453)\over s_{16} s_{24} s_{136}}\ .}}
Some aspects of the Ansatz, given in Eqs. \Ansatz, \Ansatzi\ and \AAnsatz, have also been discussed in \refs{\tye,\vaman}.
The remaining $16$ extended BCJ relations \extBCJs\ read:
$$\eqalign{
-{X_{1}\over s_{1 2} s_{5 6}}&+{X_{2}\over s_{2 3} s_{1 2 3}}+{X_{3}\over s_{1 2} s_{1 2 3}}+{X_{4}\over s_{2 3} s_{5 6}}-{X_{5}\over s_{5 6} s_{1 5 6}}-{X_{13}\over s_{5 6} s_{1 2 4}}-{X_{16}\over s_{1 5} s_{3 6}}+{X_{17}\over s_{1 2} s_{3 6}}\cr\crr
&-{X_{18}\over s_{3 6} s_{1 2 4}}-{X_{19}\over s_{1 2} s_{1 2 5}}+{X_{22}\over s_{1 5} s_{2 3}}+{X_{20}\over s_{2 3} s_{1 4 5}}+{X_{21}\over s_{3 6} s_{1 4 5}}-{X_{23}\over s_{1 5} s_{1 2 5}}-{X_{24}\over s_{1 5} s_{1 5 6}}=0\ ,\cr\cr
{X_{1}\over s_{1 2} s_{5 6}}&+{X_{5}\over s_{5 6} s_{1 5 6}}+{X_{7}\over s_{1 2} s_{1 2 6}}-{X_{8}\over s_{6 1} s_{1 2 6}}+{X_{9}\over s_{6 1} s_{1 5 6}}
-{X_{25}\over s_{12} s_{45}}-{X_{26}\over s_{4 5} s_{1 2 3}}-{X_{27}\over s_{5 6} s_{1 2 3}}\cr\crr
&-{X_{28}\over s_{4 5} s_{6 1}}+{X_{29}\over s_{1 2} s_{1 2 4}}
-{X_{30}\over s_{2 4} s_{5 6}}+{X_{31}\over s_{2 4} s_{6 1}}+{X_{32}\over s_{2 4} s_{1 2 4}}
+{X_{33}\over s_{2 4} s_{1 3 6}}+{X_{34}\over s_{4 5} s_{1 3 6}}=0\ ,\cr\cr
-{X_{7}\over s_{1 2} s_{1 2 6}}&+{X_{8}\over s_{6 1} s_{1 2 6}}+{X_{19}\over s_{1 2} s_{1 2 5}}+{X_{25}\over s_{12} s_{45}}+{X_{26}\over s_{4 5} s_{1 2 3}}+{X_{28}\over s_{4 5} s_{6 1}}-{X_{34}\over s_{4 5} s_{1 3 6}}+{X_{35}\over s_{1 2} s_{4 6}}}$$
%%%%%%%%%%%%%%%%%%%%%%%%%%%%%%%%%%%%%%%%%%%%%%%%%
$$\eqalign{
&-{X_{36}\over s_{4 6} s_{1 2 3}}-{X_{37}\over   s_{4 6} s_{1 4 6}}+{X_{38}\over s_{16} s_{1 4 6}}+{X_{39}\over s_{2 5} s_{6 1}}+{X_{40}\over s_{2 5} s_{1 2 5}}
+{X_{41}\over s_{2 5} s_{1 3 6}}-{X_{42}\over s_{2 5} s_{4 6}}=0\ ,\cr\cr
{X_{3}\over s_{1 2} s_{1 2 3}}&-{X_{19}\over s_{1 2} s_{1 2 5}}-{X_{25}\over s_{1 2} s_{4 5}}+{X_{34}\over s_{4 5} s_{1 3 6}}-{X_{40}\over s_{2 5} s_{1 2 5}}-{X_{41}\over s_{2 5} s_{1 3 6}}-{X_{43}\over s_{1 2} s_{3 4}}
+{X_{44}\over s_{4 5} s_{1 2 6}}\cr\crr
&+{X_{45}\over s_{3 4} s_{1 2 6}}
-{X_{46}\over s_{2 5} s_{3 4}}+{X_{47}\over s_{1 3} s_{2 5}}
+{X_{48}\over s_{1 3} s_{1 2 3}}-{X_{49}\over s_{1 3} s_{1 3 4}}
-{X_{50}\over s_{3 4} s_{1 3 4}}+{X_{51}\over s_{1 3} s_{4 5}}=0\ ,\cr\cr
-{X_{5}\over s_{5 6} s_{1 5 6}}&-{X_{9}\over s_{6 1} s_{1 5 6}}
+{X_{26}\over s_{4 5} s_{1 2 3}}+{X_{27}\over s_{5 6} s_{1 2 3}}+{X_{28}\over s_{4 5} s_{6 1}}+{X_{44}\over s_{4 5} s_{1 2 6}}+{X_{45}\over s_{3 4} s_{1 2 6}}
-{X_{49}\over s_{1 3} s_{1 3 4}}\cr\crr
&-{X_{50}\over s_{3 4} s_{1 3 4}}+{X_{51}\over s_{1 3} s_{4 5}}-{X_{52}\over s_{3 4} s_{5 6}}-{X_{53}\over s_{6 1} s_{1 3 6}}-{X_{54}\over s_{3 4} s_{6 1}}+{X_{55}\over s_{1 3} s_{5 6}}-{X_{56}\over s_{1 3} s_{1 3 6}}=0\ ,\cr\cr
-{X_{3}\over s_{1 2} s_{1 2 3}}&+{X_{10}\over s_{1 2} s_{3 5}}+{X_{25}\over s_{1 2} s_{4 5}}-{X_{29}\over s_{1 2} s_{1 2 4}}-{X_{32}\over s_{2 4} s_{1 2 4}}-{X_{33}\over s_{2 4} s_{1 3 6}}
-{X_{34}\over s_{4 5} s_{1 3 6}}-{X_{44}\over s_{4 5} s_{1 2 6}}\ ,\cr\crr
&-{X_{48}\over s_{1 3} s_{1 2 3}}-{X_{51}\over s_{1 3} s_{4 5}}+{X_{57}\over s_{3 5} s_{1 2 6}}-{X_{58}\over s_{2 4} s_{3 5}}+{X_{59}\over s_{1 3} s_{2 4}}-{X_{60}\over s_{1 3} s_{1 3 5}}-{X_{61}\over s_{3 5} s_{1 3 5}}=0\ ,\cr\cr
-{X_{11}\over s_{3 5} s_{6 1}}&-{X_{26}\over s_{4 5} s_{1 2 3}}-{X_{28}\over s_{4 5} s_{6 1}}+{X_{36}\over s_{4 6} s_{1 2 3}}+{X_{37}\over s_{4 6} s_{1 4 6}}
-{X_{38}\over s_{6 1} s_{1 4 6}}-{X_{44}\over s_{4 5} s_{1 2 6}}-{X_{51}\over s_{1 3} s_{4 5}}\cr\crr
&+{X_{53}\over s_{6 1} s_{1 3 6}}+{X_{56}\over s_{1 3} s_{1 3 6}}+{X_{57}\over s_{3 5} s_{1 2 6}}-{X_{60}\over s_{1 3} s_{1 3 5}}-{X_{61}\over s_{3 5} s_{1 3 5}}
-{X_{62}\over s_{3 5} s_{4 6}}+{X_{63}\over s_{1 3} s_{4 6}}=0\ ,\cr\cr
-{X_{7}\over s_{1 2} s_{1 2 6}}&+{X_{8}\over s_{6 1} s_{1 2 6}}-{X_{10}\over s_{1 2} s_{3 5}} -{X_{11}\over s_{3 5} s_{6 1}}-{X_{12}\over s_{3 5} s_{1 2 4}}-{X_{14}\over s_{3 5} s_{1 4 6}}
-{X_{17}\over s_{1 2} s_{3 6}}+{X_{18}\over s_{3 6} s_{1 2 4}}\cr\crr
&+{X_{19}\over s_{1 2} s_{1 2 5}}+{X_{39}\over s_{2 5} s_{6 1}}
+{X_{40}\over s_{2 5} s_{1 2 5}}+{X_{53}\over s_{6 1} s_{1 3 6}}
+{X_{77}\over s_{2 5} s_{3 6}}
-{X_{78}\over s_{3 6} s_{1 3 6}}-{X_{79}\over s_{2 5} s_{1 4 6}}=0\ ,\cr\cr{X_{1}\over s_{1 2} s_{5 6}}&+{X_{19}\over s_{1 2} s_{1 2 5}}-{X_{27}\over s_{5 6} s_{1 2 3}}+{X_{29}\over s_{1 2} s_{1 2 4}}+{X_{35}\over s_{1 2} s_{4 6}}-{X_{36}\over s_{4 6} s_{1 2 3}}-{X_{37}\over s_{4 6} s_{1 4 6}}
+{X_{40}\over s_{2 5} s_{1 2 5}}\cr\crr
&-{X_{42}\over s_{2 5} s_{4 6}}-{X_{66}\over s_{1 4} s_{1 2 4}}
+{X_{80}\over s_{2 5} s_{1 3 4}}-{X_{81}\over s_{1 4} s_{5 6}}
-{X_{82}\over s_{5 6} s_{1 3 4}}+{X_{83}\over s_{1 4} s_{2 5}}
+{X_{84}\over s_{1 4} s_{1 4 6}}=0\ ,\cr\cr
-{X_{3}\over s_{1 2} s_{1 2 3}}&-{X_{17}\over s_{1 2} s_{3 6}}
-{X_{29}\over s_{1 2} s_{1 2 4}}-{X_{32}\over s_{2 4} s_{1 2 4}}
-{X_{35}\over s_{1 2} s_{4 6}}
-{X_{48}\over s_{1 3} s_{1 2 3}}-{X_{56}\over s_{1 3} s_{1 3 6}}
+{X_{59}\over s_{1 3} s_{2 4}}\cr\crr
&-{X_{63}\over s_{1 3} s_{4 6}}
-{X_{72}\over s_{4 6} s_{1 3 5}}-{X_{78}\over s_{3 6} s_{1 3 6}}
-{X_{85}\over s_{2 4} s_{1 3 5}}-{X_{86}\over s_{4 6} s_{1 2 5}}
-{X_{87}\over s_{3 6} s_{1 2 5}}-{X_{88}\over s_{2 4} s_{3 6}}=0\ ,}$$
%%%%%%%%%%%%%%%%%%%%%%%%%%%%%%%%%%%%%%%%%%%%%%%%%
\eqn\extBCJs{\eqalign{
{X_{8}\over s_{6 1} s_{1 2 6}}&-{X_{11}\over s_{3 5} s_{6 1}}
-{X_{14}\over s_{3 5} s_{1 4 6}}+{X_{39}\over s_{2 5} s_{6 1}}+{X_{47}\over s_{1 3} s_{2 5}}
+{X_{53}\over s_{6 1} s_{1 3 6}}+{X_{56}\over s_{1 3} s_{1 3 6}}
-{X_{60}\over s_{1 3} s_{1 3 5}}\cr\crr
&-{X_{61}\over s_{3 5} s_{1 3 5}}+{X_{71}\over s_{2 6} s_{1 2 6}}
-{X_{79}\over s_{2 5} s_{1 4 6}}-{X_{80}\over s_{2 5} s_{1 3 4}}
+{X_{89}\over s_{2 6} s_{1 3 4}}
+{X_{90}\over s_{2 6} s_{3 5}}-{X_{91}\over s_{1 3} s_{2 6}}=0\ ,\cr\cr
{X_{2}\over s_{2 3} s_{1 2 3}}&+{X_{6}\over s_{2 3} s_{6 1}}
+{X_{31}\over s_{2 4} s_{6 1}}-{X_{37}\over s_{4 6} s_{1 4 6}}
+{X_{38}\over s_{6 1} s_{1 4 6}}-{X_{48}\over s_{1 3} s_{1 2 3}}
-{X_{53}\over s_{6 1} s_{1 3 6}}-{X_{56}\over s_{1 3} s_{1 3 6}}\cr\crr
&+{X_{59}\over s_{1 3} s_{2 4}}-{X_{63}\over s_{1 3} s_{4 6}}
-{X_{72}\over s_{4 6} s_{1 3 5}}-{X_{85}\over s_{2 4} s_{1 3 5}}
-{X_{92}\over s_{2 3} s_{4 6}}-{X_{93}\over s_{2 3} s_{1 5 6}}
-{X_{94}\over s_{2 4} s_{1 5 6}}=0\ ,\cr\cr
{X_{11}\over s_{3 5} s_{6 1}}&+{X_{12}\over s_{3 5} s_{1 2 4}}
-{X_{18}\over s_{3 6} s_{1 2 4}}+{X_{28}\over s_{4 5} s_{6 1}}
+{X_{38}\over s_{6 1} s_{1 4 6}}+{X_{44}\over s_{4 5} s_{1 2 6}}
-{X_{53}\over s_{6 1} s_{1 3 6}}-{X_{57}\over s_{3 5} s_{1 2 6}}\cr\crr
&+{X_{67}\over s_{1 4} s_{3 5}}+{X_{68}\over s_{1 4} s_{1 4 5}}
+{X_{69}\over s_{4 5} s_{1 4 5}}+{X_{78}\over s_{3 6} s_{1 3 6}}
-{X_{84}\over s_{1 4} s_{1 4 6}}-{X_{95}\over s_{1 4} s_{3 6}}
+{X_{96}\over s_{3 6} s_{4 5}}=0\ ,\cr\cr
{X_{2}\over s_{2 3} s_{1 2 3}}&+{X_{3}\over s_{1 2} s_{1 2 3}}
-{X_{10}\over s_{1 2} s_{3 5}}-{X_{14}\over s_{3 5} s_{1 4 6}}
+{X_{15}\over s_{2 3} s_{1 4 6}}-{X_{25}\over s_{1 2} s_{4 5}}
+{X_{29}\over s_{1 2} s_{1 2 4}}+{X_{44}\over s_{4 5} s_{1 2 6}}\cr\crr
&-{X_{57}\over s_{3 5} s_{1 2 6}}+{X_{64}\over s_{2 3} s_{4 5}}
+{X_{65}\over s_{1 4} s_{2 3}}-{X_{66}\over s_{1 4} s_{1 2 4}}
+{X_{67}\over s_{1 4} s_{3 5}}+{X_{68}\over s_{1 4} s_{1 4 5}}
+{X_{69}\over s_{4 5} s_{1 4 5}}=0\ ,\cr\cr
-{X_{7}\over s_{1 2} s_{1 2 6}}&+{X_{19}\over s_{1 2} s_{1 2 5}}
+{X_{23}\over s_{1 5} s_{1 2 5}}+{X_{25}\over s_{1 2} s_{4 5}}
+{X_{26}\over s_{4 5} s_{1 2 3}}+{X_{35}\over s_{1 2} s_{4 6}}
-{X_{36}\over s_{4 6} s_{1 2 3}}-{X_{69}\over s_{4 5} s_{1 4 5}}\cr\crr
&+{X_{70}\over s_{2 6} s_{4 5}}-{X_{71}\over s_{2 6} s_{1 2 6}}
+{X_{72}\over s_{4 6} s_{1 3 5}}-{X_{73}\over s_{2 6} s_{1 3 5}}
-{X_{74}\over s_{1 5} s_{4 6}}+{X_{75}\over s_{1 5} s_{2 6}}
-{X_{76}\over s_{1 5} s_{1 4 5}}=0\ ,\cr\cr
{X_{2}\over s_{2 3} s_{1 2 3}}&+{X_{3}\over s_{1 2} s_{1 2 3}}
+{X_{6}\over s_{2 3} s_{6 1}}-{X_{7}\over s_{1 2} s_{1 2 6}}
+{X_{8}\over s_{6 1} s_{1 2 6}}+{X_{35}\over s_{1 2} s_{4 6}}
-{X_{37}\over s_{4 6} s_{1 4 6}}+{X_{38}\over s_{6 1} s_{1 4 6}}\cr\crr
&-{X_{43}\over s_{1 2} s_{3 4}} +{X_{54}\over s_{3 4} s_{6 1}}
+{X_{86}\over s_{4 6} s_{1 2 5}}-{X_{92}\over s_{2 3} s_{4 6}}
-{X_{93}\over s_{2 3} s_{1 5 6}}+{X_{97}\over s_{3 4} s_{1 2 5}}
-{X_{98}\over s_{3 4} s_{2 3 4}}=0\ .}}
Above, the one hundred  numerator triplets $X_i$ are defined as follows:
$$\eqalign{
X_{1}&=n_{1}(2 3 4 5)-n_{4}(2 3 4 5)+n_{4}(2 4 3 5)\ \ \ ,\ \ \  
X_{2}=n_{2}(2 3 4 5)-n_{5}(2 3 4 5)+n_{5}(2 3 5 4) \ ,\cr
X_{3}&=n_{3}(2 3 4 5)-n_{4}(2 3 4 5)+n_{4}(2 3 5 4) \ \ \ ,\ \ \ 
X_{4}=-n_{1}(4 2 3 5)+n_{5}(2 3 4 5)-n_{6}(2 3 4 5) \ ,\cr
X_{5}&=-n_{6}(2 3 4 5)+n_{6}(2 4 3 5)+n_{7}(2 3 4 5) \ \ \ ,\ \ \ 
X_{6}=n'_1(2 3 4 5)-n'_7(2 3 4 5)+n'_7(2 3 5 4) \ ,\cr
X_{7}&=n'_2(2 3 4 5)-n'_2(2 3 5 4)-n'_3(2 3 4 5) \ \ \ ,\ \ \ 
X_{8}=n'_4(2 3 4 5)-n'_5(2 3 4 5)+n'_5(2 3 5 4) \ ,\cr}$$
$$\eqalign{
X_{9}&=n'_6(2 3 4 5)-n'_7(2 3 4 5)+n'_7(2 4 3 5)\ \ \ ,\ \ \ 
X_{10}=-n_{1}(2 3 5 4)+n_{3}(2 4 3 5)+n'_2(2 3 5 4) \ ,\cr
X_{11}&=n'_1(2 4 3 5)+n'_5(2 3 5 4)-n'_6(2 3 5 4) \ \ \ ,\ \ \ 
X_{12}=n_{2}(2 4 3 5)-n_{3}(2 4 3 5)+n_{3}(4 2 3 5) \ ,\cr
X_{13}&=-n_{4}(2 4 3 5)+n_{4}(4 2 3 5)+n_{5}(2 4 3 5) \ \ \ ,\ \ \ 
X_{14}=-n_{7}(2 3 5 4)+n'_3(4 2 3 5)+n'_6(2 3 5 4) \ ,\cr
X_{15}&=n_{6}(2 3 5 4)-n'_2(4 2 3 5)-n'_7(2 3 5 4) \ \ \ ,\ \ \ 
X_{16}=n_{1}(5 2 4 3)-n_{4}(5 2 4 3)+n_{4}(5 4 2 3) \ ,\cr
X_{17}&=n_{1}(2 4 5 3)-n_{4}(2 4 5 3)+n_{4}(2 5 4 3) \ \ \ ,\ \ \ 
X_{18}=-n_{4}(2 4 5 3)+n_{4}(4 2 5 3)+n_{5}(2 4 5 3) \ ,\cr
X_{19}&=n_{3}(2 5 3 4)-n_{4}(2 5 3 4)+n_{4}(2 5 4 3) \ \ \ ,\ \ \ 
X_{20}=n_{2}(4 5 2 3)-n_{3}(4 5 2 3)+n_{3}(5 4 2 3) \ ,\cr
X_{21}&=-n_{4}(4 5 2 3)+n_{4}(5 4 2 3)+n_{5}(4 5 2 3) \ \ \ ,\ \ \ 
X_{22}=n_{1}(5 2 3 4)-n_{3}(5 4 2 3)-n'_2(5 2 3 4) \ ,\cr
X_{23}&=n_{3}(5 2 3 4)-n_{4}(5 2 3 4)+n_{4}(5 2 4 3) \ \ \ ,\ \ \ 
X_{24}=-n'_2(5 2 3 4)+n'_2(5 2 4 3)+n'_3(5 2 3 4) \ ,\cr
X_{25}&=-n_{1}(2 4 5 3)+n_{3}(2 3 4 5)-n'_3(2 3 4 5) \ \ \ ,\ \ \ 
X_{26}=n_{2}(2 3 4 5)-n_{3}(2 3 4 5)+n_{3}(3 2 4 5)\ ,\cr
X_{27}&=-n_{4}(2 3 4 5)+n_{4}(3 2 4 5)+n_{5}(2 3 4 5) \ \ \ ,\ \ \ 
X_{28}=n'_1(2 3 4 5)-n'_4(2 3 4 5)-n'_6(2 4 5 3) \ ,\cr
X_{29}&=n_{3}(2 4 3 5)-n_{4}(2 4 3 5)+n_{4}(2 4 5 3) \ \ \ ,\ \ \ 
X_{30}=n_{1}(3 2 4 5)-n_{5}(2 4 3 5)+n_{6}(2 4 3 5) \ ,\cr
X_{31}&=n'_1(2 4 3 5)-n'_7(2 4 3 5)+n'_7(2 4 5 3) \ \ \ ,\ \ \ 
X_{32}=n_{2}(2 4 3 5)-n_{5}(2 4 3 5)+n_{5}(2 4 5 3) \ ,\cr
X_{33}&=n_{6}(2 4 5 3)-n'_2(3 2 4 5)-n'_7(2 4 5 3) \ \ \ ,\ \ \ 
X_{34}=n_{7}(2 4 5 3)-n'_3(3 2 4 5)-n'_6(2 4 5 3)  \ ,\cr
X_{35}&=n_{1}(2 3 5 4)-n_{4}(2 3 5 4)+n_{4}(2 5 3 4) \ \ \ ,\ \ \ 
X_{36}=-n_{4}(2 3 5 4)+n_{4}(3 2 5 4)+n_{5}(2 3 5 4)\ ,\cr
X_{37}&=n_{6}(2 3 5 4)-n_{6}(2 5 3 4)-n_{7}(2 3 5 4) \ \ \ ,\ \ \ 
X_{38}=n'_6(2 3 5 4)-n'_7(2 3 5 4)+n'_7(2 5 3 4) \ ,\cr
X_{39}&=n'_1(2 5 3 4)-n'_7(2 5 3 4)+n'_7(2 5 4 3) \ \ \ ,\ \ \ 
X_{40}=n_{2}(2 5 3 4)-n_{5}(2 5 3 4)+n_{5}(2 5 4 3) \ ,\cr
X_{41}&=n_{6}(2 5 4 3)-n'_2(3 2 5 4)-n'_7(2 5 4 3) \ \ \ ,\ \ \ 
X_{42}=n_{1}(3 2 5 4)-n_{5}(2 5 3 4)+n_{6}(2 5 3 4)\ ,\cr
X_{43}&=n_{1}(2 3 4 5)-n_{3}(2 5 3 4)-n'_2(2 3 4 5)  \ \ \ ,\ \ \
X_{44}=n_{7}(3 4 5 2)-n'_3(2 3 4 5)+n'_4(2 3 4 5) \ ,\cr
X_{45}&=n_{6}(3 4 5 2)-n'_2(2 3 4 5)+n'_5(2 3 4 5)  \ \ \ ,\ \ \
X_{46}=-n_{2}(2 5 3 4)+n_{2}(3 4 2 5)+n'_1(2 5 3 4) \ ,\cr
X_{47}&=n_{1}(3 2 5 4)-n_{3}(3 4 2 5)-n'_2(3 2 5 4)  \ \ \ ,\ \ \
X_{48}=n_{3}(3 2 4 5)-n_{4}(3 2 4 5)+n_{4}(3 2 5 4) \ ,\cr
X_{49}&=-n_{3}(3 4 2 5)+n_{4}(3 4 2 5)-n_{4}(3 4 5 2)  \ \ \ ,\ \ \
X_{50}=-n_{2}(3 4 2 5)+n_{5}(3 4 2 5)-n_{5}(3 4 5 2) \ ,\cr
X_{51}&=n_{1}(3 4 5 2)-n_{3}(3 2 4 5)+n'_3(3 2 4 5)  \ \ \ ,\ \ \
X_{52}=n_{1}(2 3 4 5)-n_{5}(3 4 2 5)-n_{7}(2 3 4 5) \ ,\cr
X_{53}&=-n'_6(2 4 5 3)+n'_7(2 4 5 3)-n'_7(2 5 4 3)  \ \ \ ,\ \ \
X_{54}=n'_1(2 5 3 4)+n'_5(2 3 4 5)-n'_6(2 3 4 5) \ ,\cr
X_{55}&=n_{1}(3 2 4 5)-n_{4}(3 2 4 5)+n_{4}(3 4 2 5)  \ \ \ ,\ \ \
X_{56}=-n'_2(3 2 4 5)+n'_2(3 2 5 4)+n'_3(3 2 4 5) \ ,\cr
X_{57}&=n_{6}(3 5 4 2)-n'_2(2 3 5 4)+n'_5(2 3 5 4)  \ \ \ ,\ \ \
X_{58}=-n_{2}(2 4 3 5)+n_{2}(3 5 2 4)+n'_1(2 4 3 5) \ ,\cr
X_{59}&=n_{1}(3 2 4 5)-n_{3}(3 5 2 4)-n'_2(3 2 4 5)  \ \ \ ,\ \ \
X_{60}=-n_{3}(3 5 2 4)+n_{4}(3 5 2 4)-n_{4}(3 5 4 2) \ ,\cr
X_{61}&=-n_{2}(3 5 2 4)+n_{5}(3 5 2 4)-n_{5}(3 5 4 2)  \ \ \ ,\ \ \
X_{62}=n_{1}(2 3 5 4)-n_{5}(3 5 2 4)-n_{7}(2 3 5 4) \ ,\cr
X_{63}&=n_{1}(3 2 5 4)-n_{4}(3 2 5 4)+n_{4}(3 5 2 4)  \ \ \ ,\ \ \
X_{64}=-n_{2}(2 3 4 5)+n_{2}(4 5 2 3)+n'_1(2 3 4 5) \ ,\cr
X_{65}&=-n_{1}(4 2 3 5)+n_{3}(4 5 2 3)+n'_2(4 2 3 5)  \ \ \ ,\ \ \
X_{66}=-n_{3}(4 2 3 5)+n_{4}(4 2 3 5)-n_{4}(4 2 5 3) \ ,\cr}$$
%%%%%%%%%%%%%%%%%%%%%%%%%%%%%%%%%%%%%%%%%%%%%%%%%
\eqn\NNs{\eqalign{
X_{67}&=n_{1}(4 3 5 2)-n_{3}(4 2 3 5)+n'_3(4 2 3 5)  \ \ \ ,\ \ \
X_{68}=-n_{3}(4 5 2 3)+n_{4}(4 5 2 3)-n_{4}(4 5 3 2) \ ,\cr
X_{69}&=-n_{2}(4 5 2 3)+n_{5}(4 5 2 3)-n_{5}(4 5 3 2)  \ \ \ ,\ \ \
X_{70}=n_{1}(3 4 5 2)-n_{5}(4 5 3 2)-n_{7}(3 4 5 2) \ ,\cr
X_{71}&=n_{6}(3 4 5 2)-n_{6}(3 5 4 2)-n_{7}(3 4 5 2)  \ \ \ ,\ \ \
X_{72}=-n_{4}(3 5 2 4)+n_{4}(5 3 2 4)+n_{5}(3 5 2 4)\ ,\cr
X_{73}&=n_{4}(3 5 4 2)-n_{4}(5 3 4 2)-n_{5}(3 5 4 2)  \ \ \ ,\ \ \
X_{74}=n_{1}(5 2 3 4)-n_{4}(5 2 3 4)+n_{4}(5 3 2 4) \ ,\cr
X_{75}&=n_{1}(5 3 4 2)-n_{4}(5 3 4 2)+n_{4}(5 4 3 2)  \ \ \ ,\ \ \
X_{76}=n_{3}(5 4 2 3)-n_{4}(5 4 2 3)+n_{4}(5 4 3 2) \ ,\cr
X_{77}&=n_{1}(4 2 5 3)-n_{5}(2 5 4 3)+n_{6}(2 5 4 3)  \ \ \ ,\ \ \
X_{78}=-n_{6}(2 4 5 3)+n_{6}(2 5 4 3)+n_{7}(2 4 5 3) \ ,\cr
X_{79}&=n_{6}(2 5 3 4)-n'_2(4 2 5 3)-n'_7(2 5 3 4)  \ \ \ ,\ \ \
X_{80}=n_{2}(3 4 2 5)-n_{3}(3 4 2 5)+n_{3}(4 3 2 5) \ ,\cr
X_{81}&=n_{1}(4 2 3 5)-n_{4}(4 2 3 5)+n_{4}(4 3 2 5)  \ \ \ ,\ \ \
X_{82}=n_{4}(3 4 2 5)-n_{4}(4 3 2 5)-n_{5}(3 4 2 5) \ ,\cr
X_{83}&=n_{1}(4 2 5 3)-n_{3}(4 3 2 5)-n'_2(4 2 5 3)  \ \ \ ,\ \ \
X_{84}=-n'_2(4 2 3 5)+n'_2(4 2 5 3)+n'_3(4 2 3 5) \ ,\cr
X_{85}&=n_{2}(3 5 2 4)-n_{3}(3 5 2 4)+n_{3}(5 3 2 4)  \ \ \ ,\ \ \
X_{86}=-n_{4}(2 5 3 4)+n_{4}(5 2 3 4)+n_{5}(2 5 3 4) \ ,\cr
X_{87}&=-n_{4}(2 5 4 3)+n_{4}(5 2 4 3)+n_{5}(2 5 4 3)  \ \ \ ,\ \ \
X_{88}=n_{1}(5 2 4 3)-n_{5}(2 4 5 3)+n_{6}(2 4 5 3) \ ,\cr
X_{89}&=-n_{4}(3 4 5 2)+n_{4}(4 3 5 2)+n_{5}(3 4 5 2)  \ \ \ ,\ \ \
X_{90}=n_{1}(4 3 5 2)-n_{5}(3 5 4 2)+n_{6}(3 5 4 2) \ ,\cr
X_{91}&=n_{1}(3 4 5 2)-n_{4}(3 4 5 2)+n_{4}(3 5 4 2) \ \ \ ,\ \ \
X_{92}=-n_{1}(5 2 3 4)+n_{5}(2 3 5 4)-n_{6}(2 3 5 4)\ ,\cr
X_{93}&=n_{6}(2 3 4 5)-n'_2(5 2 3 4)-n'_7(2 3 4 5)  \ \ \ ,\ \ \
X_{94}=n_{6}(2 4 3 5)-n'_2(5 2 4 3)-n'_7(2 4 3 5) \ ,\cr
X_{95}&=n_{1}(4 2 5 3)-n_{4}(4 2 5 3)+n_{4}(4 5 2 3)  \ \ \ ,\ \ \
X_{96}=n_{1}(2 4 5 3)-n_{5}(4 5 2 3)-n_{7}(2 4 5 3) \ ,\cr
X_{97}&=n_{2}(2534)-n_{3}(2534)+n_3(5234)\ \ \ ,\ \ \ 
X_{98}=n_7(2345)-n'_3(5234)-n'_6(2345)\ ,\cr
X_{99}&=n_4(4532)-n_4(5432)-n_5(4532)\ \ \ ,\ \ \ 
X_{100}=-n_1(5342)+n_5(3452)-n_6(3452)\ .}}  
Finally, one can check that the solution to the equations
$X_i=0$ gives rise to $81$ independent
relations between the $105$ numerators $n_i$ such that they can
be written in terms of 
the $24$ independent numerators \setkin:
\eqn\numtwnty{\eqalign{
m_{1}&=n_{3}(2345)\quad ,\quad
m_{2}=n_{3}(2435)\quad ,\quad
m_{3}=n_{3}(3245)\quad ,\quad
m_{4}=n_{3}(3425)\quad \cr
m_{5}&=n_{3}(4235)\quad ,\quad
m_{6}=n_{3}(4325)\quad ,\quad
m_{7}=n_{4}(2345)\quad ,\quad
m_{8}=n_{4}(2435)\quad \cr
m_{9}&=n_{4}(3245)\quad ,\quad
m_{10}=n_{4}(3425)\ \ ,\quad
m_{11}=n_{4}(4235)\ \ ,\quad
m_{12}=n_{4}(4325)\quad \cr
m_{13}&=n'_2(2354)=-n'_3(2435)\quad ,\quad
m_{14}=n'_2(3254)=-n'_3(3425)\quad ,\cr
m_{15}&=n'_2(4253)=-n'_3(4325)\quad ,\quad
m_{16}=n'_3(2345)\ \ \ ,\ \ \ m_{17}=n'_3(3245)\quad\cr
m_{18}&=n'_3(4235)\quad ,\quad
m_{19}=n'_4(2345)\quad ,\quad
m_{20}=n'_5(2354)=-n'_4(2435)\ ,\cr
m_{21}&=n'_6(2354)=-n'_4(4235)\quad ,\quad
m_{22}=n'_6(2453)=-n'_4(3245)\ ,\cr
m_{23}&=n'_7(2435)=n'_4(3425)\quad\ \ ,\quad\
m_{24}=n'_7(2534)=n'_4(4325)\ .}}
The equalities in \numtwnty\ relating two different kinematic factors $n'_i$ and $n'_j$ follow
from the symmetry properties of the BRST building blocks $T_{ijk}$ and $T_{ijkl}$. For example,
to prove the identity $n'_2(2354)=-n'_3(2435)$ one applies the parity transformation
$1\leftrightarrow 5, 2\leftrightarrow 4$ in the kinematic factors of \ndefs\ to get
$n'_2(2345) = \langle (T_{534} - T_{543})T_{21}V_6\rangle$ and $n'_3(2345) =
- \langle T_{543}T_{21}V_6\rangle$. Therefore
%\eqn\trueone{
$
n'_2(2354) = \langle (T_{435} - T_{453})T_{21}V_6\rangle = \langle T_{534}T_{21}V_6\rangle = - n'_3(2435),
$
%}
where the second equality follows from $T_{\{ijk\}} = T_{(ij)k} =0$. 
To prove $n'_6(2354)=-n'_4(4235)$ first use the properties 
$T_{ijkl} = T_{[ij]kl}$, $T_{[ijk]l}= T_{ij[kl]} + T_{kl[ij]} = 0$ to rewrite
\eqn\nsixhook{
n_6(2345) = \langle (T_{1234} - T_{1324} - T_{1423} + T_{1432})V_5V_6\rangle = - \langle T_{2341}V_5V_6\rangle
}
which implies under parity that $n'_6(2345) = -\langle T_{4325}V_1V_6\rangle$. Furthermore, the parity transformation
of $n_4(2345)$ in \ndefs\ results in $n'_4(2345) = \langle T_{5432}V_1V_6\rangle$ and therefore one finally obtains
%\eqn\truetwo{
$
n'_6(2354) = -\langle T_{5324}V_1V_6\rangle = - n'_4(4235).
$
%}
The other identities in \numtwnty\ are easily shown using similar manipulations.

\appendix{D}{The three--vertex field--theory diagrams}

Finally  in this appendix we draw all $14$ diagrams involving only 
three--vertices and their corresponding pure spinor superspace expressions.
\vskip1cm
\centerline{\epsfxsize 5truein\epsfbox{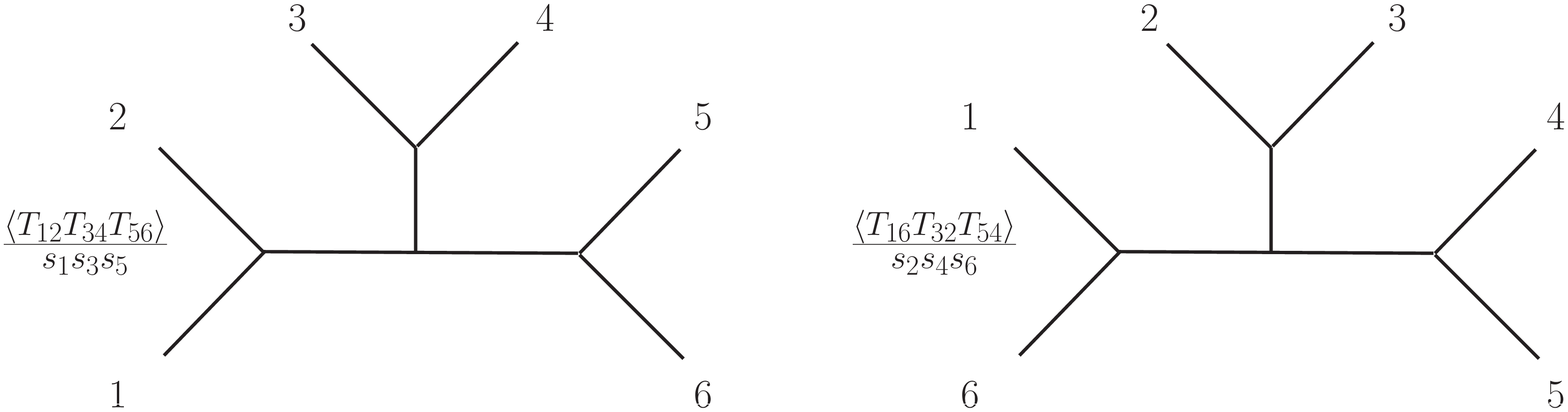}}
%\smallskip
\centerline{\epsfxsize 5truein\epsfbox{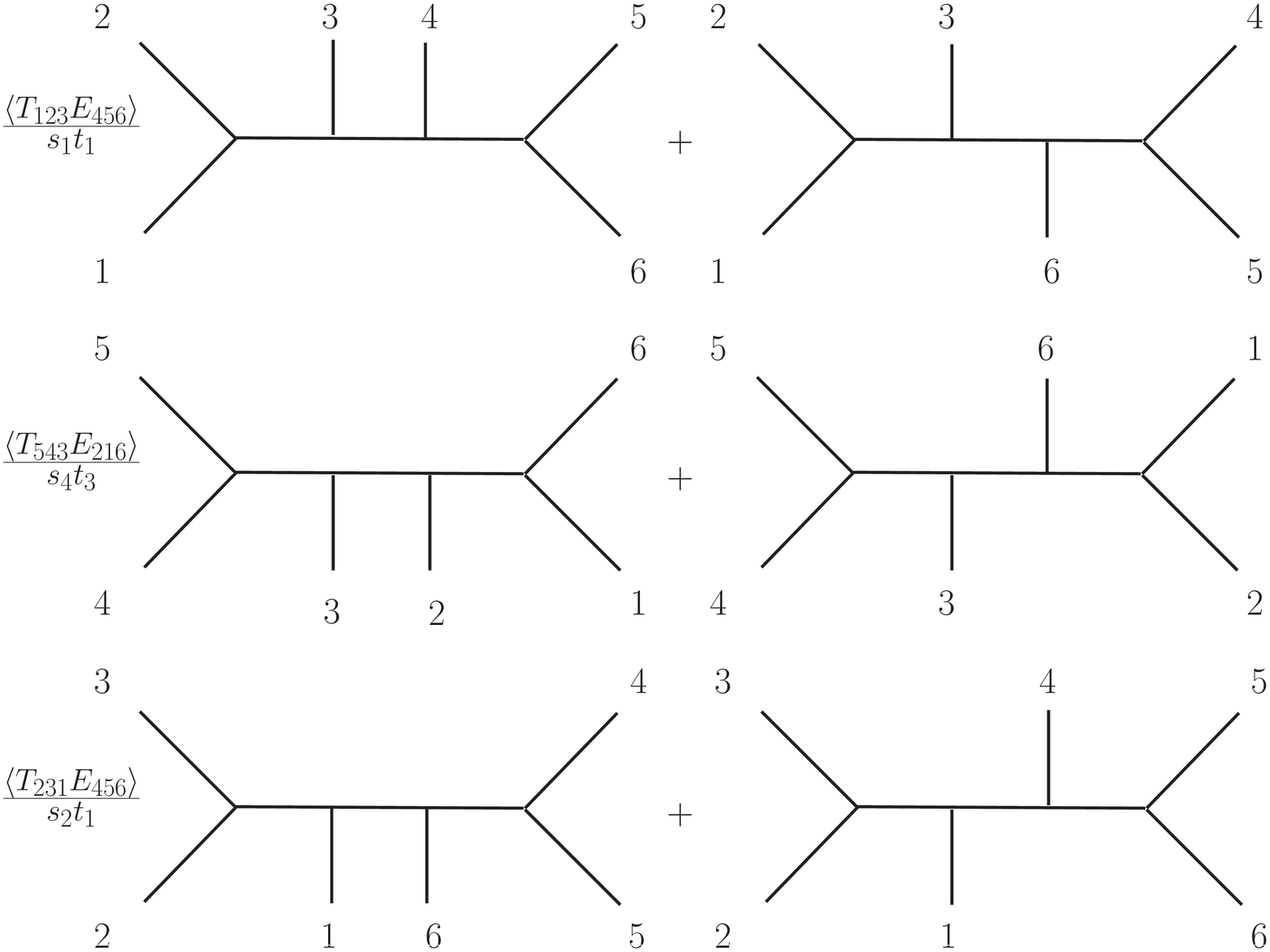}}
%\smallskip
\vskip1cm
\centerline{\epsfxsize 5truein\epsfbox{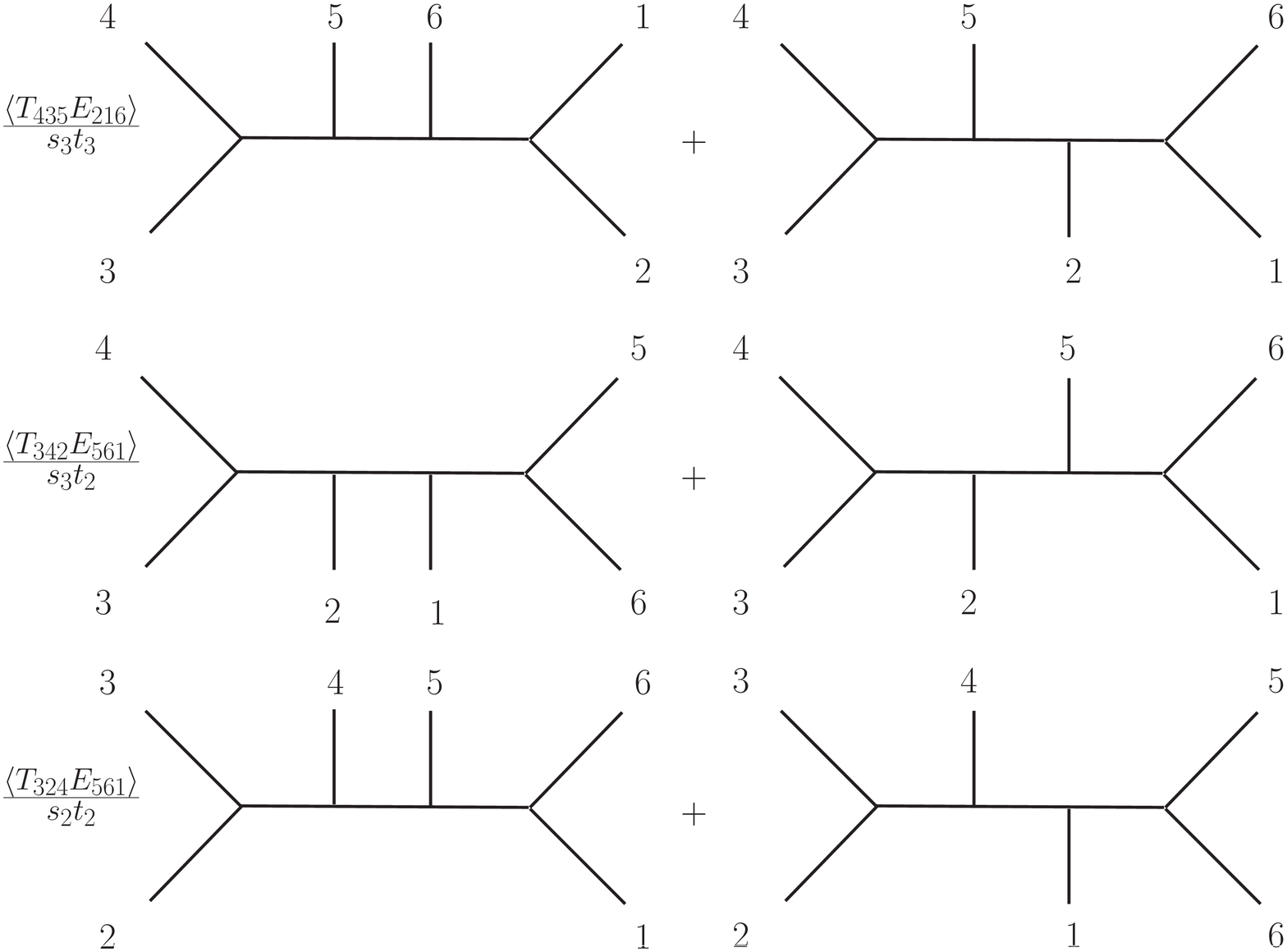}}
\vskip0.75cm
\centerline{\ninepoint \baselineskip=2pt {\bf Fig. 4.} 
{The $14$ field theory diagrams and their corresponding
pure spinor superspace expressions.}}
\smallskip
\bigskip

\listrefs

\end